\documentclass[aps,superscriptaddress,twocolumn,twoside,floatfix,pra,nofootinbib,a4paper]{revtex4-2}

\usepackage{times}
\usepackage{amsfonts}
\usepackage{amsmath}
\usepackage{amssymb}
\usepackage{amsthm}
\usepackage{multirow}
\usepackage[normalem]{ulem}
\usepackage[T1]{fontenc}
\usepackage{float}
\usepackage{mathtools}
\usepackage{bm}
\usepackage[UKenglish,cleanlook]{isodate}   
\usepackage{mathrsfs}
\usepackage{dsfont}

\usepackage{physics}

\usepackage{xcolor}
\definecolor{carmine}{RGB}{150,0,24}
\definecolor{mycolor1}{rgb}{1.00000,0.00000,1.00000}

\usepackage{natbib}
\usepackage[colorlinks=true,linkcolor=blue,citecolor=magenta,urlcolor=blue]{hyperref}

\usepackage{graphicx}
\usepackage{tikz-cd}

\begin{document}


\title{Maximal steady-state entanglement in autonomous quantum thermal machines}


\author{Shishir Khandelwal*}
\affiliation{Physics Department and NanoLund, Lund University, Box 118, 22100 Lund, Sweden}
\email{shishir.khandelwal@teorfys.lu.se}

\author{Björn Annby-Andersson}
\affiliation{Physics Department and NanoLund, Lund University, Box 118, 22100 Lund, Sweden}

\author{Giovanni Francesco Diotallevi}
\affiliation{Physics Department and NanoLund, Lund University, Box 118, 22100 Lund, Sweden}
\affiliation{Augsburg University$,$ Institute of Physics$,$ Universitätsstraße 1 (Physik Nord)$,$ 86159 Augsburg}

\author{Andreas Wacker}
\affiliation{Physics Department and NanoLund, Lund University, Box 118, 22100 Lund, Sweden}

\author{Armin Tavakoli}
\affiliation{Physics Department and NanoLund, Lund University, Box 118, 22100 Lund, Sweden}

\begin{abstract}
We devise an autonomous quantum thermal machine  consisting of three pairwise-interacting qubits, two of which are locally coupled to thermal reservoirs. The machine operates autonomously, as it requires no time-coherent control, external driving or quantum bath engineering, and is instead propelled by a chemical potential bias. Under ideal conditions, we show that this out-of-equilibrium system can deterministically generate a maximally entangled steady-state between two of the qubits, or any desired pure two-qubit entangled state, emerging as a dark state of the system. We study the robustness of entanglement production with respect to several relevant parameters, obtaining nearly-maximally-entangled states well-away from the ideal regime of operation. Furthermore, we show that our machine architecture can be generalised to a configuration with $2n-1$ qubits, in which only a potential bias and  two-body interactions are sufficient to generate genuine multipartite maximally entangled steady states in the form of a W state of $n$ qubits.  

\end{abstract}


\maketitle


\section{Introduction}
  Quantum thermal machines are open systems of interacting quanta that harvest spontaneous interactions with thermal reservoirs to perform a designated task. These machines  have been proposed as quantum mechanical counterparts to the classical thermal machines of the industrial age, for instance, for  work extraction, heating, cooling and keeping time \cite{Scovil1959,Alicki1979,Kosloff2014, Mitchison2019, Bhattacharjee2021}. In recent years, they have also been implemented in experiments~\cite{Rosnagel2014, Rosnagel2016, Maslennikov2019, Klatzow2019, Pearson2021}.   However, quantum thermal machines can go further, and perform tasks that themselves are inherently quantum mechanical. The paradigmatic example is the generation of entanglement. It is well-known that entanglement can be generated via dissipation, and the topic has received much  interest \cite{Diehl2008, Verstraete2009, Kastoryano2011, Krauter2011, Lin2013, Shankar2013}. When operating out-of-equilibrium, it is achieved by external driving of the system \cite{Mancilla2009,Rao2013,Khandelwal2023} or engineering of the quantum reservoirs \cite{Banerjee2010, Reiter2012, Tacchino2018}.

This progress spurred the question of identifying the minimal resources to generate steady-state entanglement in dissipative out-of-equilibrium systems, i.e.~with a time-independent Hamiltonian, no external work sources and no quantum bath engineering. The challenge is to rely only on spontaneous interactions with an uncontrolled environment while the system is not in equilibrium. Interestingly, an affirmative answer has been given. Two interacting qubits that are individually coupled to reservoirs of different temperature can end up in an entangled steady-state \cite{Brask2015}, due to a heat current  through the system  \cite{Khandelwal2020}. Unfortunately, however, the generated entanglement is weak and unable to defy several notions of classicality \cite{Brask2022operational}. Going beyond the two-qubit systems, maximal entanglement is possible by using networks of several qubits \cite{Poulsen2022, Poulsen2022b}. However, in aiming for a minimal machine that produces maximal entanglement, several works have considered supplementing the two-qubit system with some additional, non-autonomous, resources  to amplify the two-qubit entanglement \cite{Tacchino2018, Brask2022operational, Tavakoli2018, Tavakoli2020, Diotallevi2023,Oliveira2024}. Amplification of the entanglement has been possible also in the fully autonomous setting \cite{Man2019}, in particular by leveraging a voltage bias instead of a temperature bias \cite{Prech2023}. Nevertheless, none of these approaches have been able to deterministically generate maximal, or even nearly maximal, steady-state entanglement. 

\begin{figure}
	\centering
	\includegraphics[width=0.8\columnwidth]{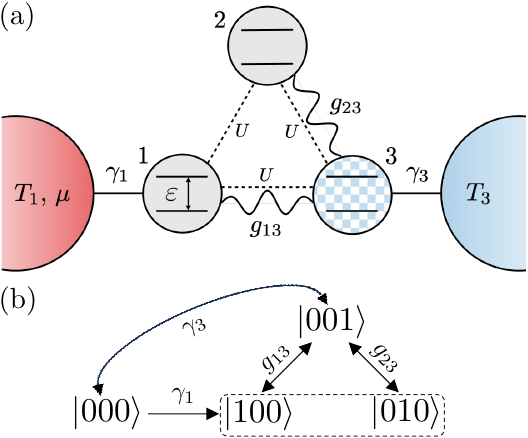}
	\caption{(a) The three-qubit autonomous thermal machine. The chemical potential of the left reservoir is the key resource for entanglement generation. The steady-state for the ideal regime of operation,  has the solid gray qubits in the state $\ket{\Psi^-}$ while the shaded-blue qubit in the state $\ket{0}$.  (b) A representative flow chart showing the relevant states of the three qubits. The states $\ket{100}$, $\ket{010}$ and $\ket{001}$ effectively form a lambda system \cite{Lambropoulos2007}. The dashed box indicates the entangled subspace corresponding to the steady-state.
	}
	\label{fig:setup}
\end{figure}

In this work, we identify the minimal thermal machine for generating any pure entangled state between the degrees of freedom of two physically separated qubits, using only time-independent, i.e.~autonomous, resources.  The machine exploits a chemical potential bias between two fermionic reservoirs at equilibrium, and involves three qubits subject to Coulomb repulsion and pairwise flip-flop interaction. Our machine, illustrated in Figure~\ref{fig:setup}(a), thus uses the third qubit as a mediator for entanglement generation. Furthermore, we show that the machine performs well beyond the ideal regime of operation; reasonable Coulomb forces and potential biases, small detunings, and coupling strength variations all lead to nearly-maximal steady-state entanglement.  Motivated by this, we investigate the conceptually deeper question of whether autonomous machines based only on two-body Hamiltonians are able to generate maximal entanglement between \textit{any} number of qubits. We answer this in the positive by identifying a  generalisation of our machine, which features $2n-1$ qubits, and prove that it can render $n$ of these qubits in a W-type steady-state.

\section{Results}

\subsection{Model}
We consider a setup of three interacting qubits, as shown in Fig.~\ref{fig:setup}(a). The excited state of each qubit corresponds to the energy gap $\varepsilon$. Qubit pairs 1-3 and 2-3 interact via flip-flop interactions with coupling strengths $g_{13}$ and $g_{23}$ respectively. Every qubit pair also interacts electrostatically via a Coulomb repulsion of magnitude $U$. The Hamiltonian of the three qubits is therefore given by ($\hbar=k_B=1$),
\begin{equation}
\begin{aligned}\label{eq:ham}
	H &= \sum_{i=1}^{3} \varepsilon \sigma_+^{(i)}\sigma_-^{(i)} +\hspace*{-0.2cm} \sum_{j=\{1,2\}} \left\{g_{j3}\left(\sigma^{(j)}_+\sigma^{(3)}_- +  \sigma^{(j)}_-\sigma^{(3)}_+\right)\right\} \\
	&+U\left( \ketbra{110}{110}+\ketbra{101}{101}+\ketbra{011}{011} + 3\ketbra{111}{111}\right)
\end{aligned}
\end{equation}where $\sigma_{+}^{(j)}\,(\sigma_{-}^{(j)})$ is the raising (lowering operator) for qubit $j$.
 {Such a model can be realized in quantum dot systems \cite{ReimannRMP2002} with  spin-polarized electrons. For example, in Refs.~\cite{GaudreauPRL2006,RoggeNJP2009} such a triangular system was studied in a two-dimensional electron gas. The authors managed  to bring all excitation energies  into resonance \cite{RoggeNJP2009} and  
suitable values of Coulomb interactions and tunnel coupling could be achieved \cite{GaudreauPRL2006}. Alternatives could be based on pilar structures \cite{AmahaPhysicaE2008} or graphene bilayers \cite{BischoffNJP2013}, where such triple dots have been realized. A further interesting approach could apply} ultracold fermionic atoms \cite{BrantutScience2012}. Similar to the transport experiments \cite{GaudreauPRL2006,RoggeNJP2009}, we take qubits 1 and 3 to be coupled to equilibrated reservoirs of non-interacting fermions, with bare coupling strengths $\gamma_1$ and $\gamma_3$. Throughout this article, we restrict ourselves to the regime of operation in which $\gamma_j\ll\text{max}\{T_j,\lvert\varepsilon\pm g_{ij}-\mu_j\rvert\}$, where $T_j$ and $\mu_j$ are the temperature and chemical potential respectively, of reservoir $j$. Then, the evolution of the system can accurately be modelled by a Lindblad master equation. Further imposing $g_{ij}\ll \text{max}\{T_j,\lvert\varepsilon-\mu_j\rvert\}$ ensures that the dissipation acts locally on the qubits \cite{Potts2021,Hofer2017}. Therefore, the Lindblad equation takes the following form,
\begin{equation}
\begin{aligned}\label{eq:lind}
	\dot\rho =\mathcal L \rho= -i\left[ H,\rho\right] +\hspace*{-0.4cm} \sum_{\substack{j\in\{1,3\}\\ p,q\in\{0,1\}}}\hspace*{-0.3cm} \left(\gamma_{jpq}^+ \mathcal D\left[ L_{jpq}\right]\rho + \gamma_{jpq}^-\mathcal D\big[ L^\dagger_{jpq}\big]\rho\right),
\end{aligned}
\end{equation}where the dissipators are $\mathcal D[L]\rho\coloneqq L\rho L^\dagger - \frac{1}{2}\left( L^\dagger L\rho + \rho L^\dagger L\right)$ and $\gamma_{jpq}^{+}$ and $\gamma_{jpq}^{-}$ are the rates corresponding to transitions induced by the Lindblad jump operators $L_{jpq}$ and $L_{jpq}^\dagger$. These are respectively defined as, $L_{1pq}\coloneqq\ketbra{1pq}{0pq}$ and $L_{3pq}\coloneqq\ketbra{pq1}{pq0}$. The exact excitation and relaxation rates are determined by the statistics of the reservoirs. For our reservoirs, we have that $\gamma^+_{jpq} = \gamma_j n_F\left(\varepsilon+\mathcal U_{pq},\mu_j,T_j\right)$ and $\gamma^-_{jpq}= \gamma_j \left( 1-n_F\left(\varepsilon+\mathcal U_{pq},\mu_j,T_j\right)\right)$,  where $n_F\left(\varepsilon,\mu,T \right)=1/(e^{(\varepsilon-\mu)/T}+1)$ is the Fermi-Dirac distribution. $\mathcal U_{pq} \coloneqq U\delta_{p+q,1}+2U\delta_{p+q,2}$ takes into account the Coulomb interaction. It ensures that the potential energy difference of $U$ is added for each additional excitation created in the system; $U$ for a second excitation and $2U$ for a third. We take $\mu_1=\mu$ and for simplicity, we set $\mu_3=0$ \cite{footnote}. Therefore, the system is driven by two reservoirs which are out-of-equilibrium with each other. In the case of equal temperatures of the reservoirs, the imbalance is determined solely by the chemical potential of the left reservoir.

\subsection{Generation of arbitrary pure entangled states}
 We now show that there exists an ideal parameter regime in which the steady-state solution to Eq.~\eqref{eq:lind}, when reduced to qubits 1 and 2, can correspond to any pure entangled state (up to local unitaries). As in Ref.~\cite{Prech2023}, we consider the limit in which 
\begin{align}\label{eq:lim}
U\to\infty \quad \text{and} \quad\mu\to\infty,\,\,\text{with}\,\, \quad U/\mu\to\infty.
\end{align}The first limit ensures that whenever there is already an excitation in the system, the fermions in either reservoir cannot overcome the large Coulomb interaction to excite the system further. On the level of rates, this translates to $\gamma_{jpq}^+=0$ whenever $p+q\geq1$. This eliminates the possibility of double or triple excitation in the three-qubit steady state. The second limit ensures that when there is no excitation in the system, the population in reservoir 1 is filled, i.e., $n_F(\varepsilon,\mu,T_1)=1$. The last limit is important to ensure that no double or triple excitations are possible at any point in the evolution of the machine. The reservoirs are therefore, extremely out-of-equilibrium with each other, with the left reservoir (coupled with qubit 1) at an extremely high bias and the right reservoir (coupled with qubit 3) at a low or zero bias.  The only pathway for an excitation to leave the system is through qubit 3 and the right reservoir. We therefore refer to this qubit as the sink qubit. In the limits \eqref{eq:lim}, the only relevant transitions are induced by the jump operators $L_{100}$, $L_{300}$ and $L_{300}^\dagger$ \cite{footnote2}. A matrix form of the Liouvillian $\mathcal L$ in these limits  can be found in the Supplementary Information. The steady-state of the machine is the unique eigenstate of $\mathcal L$ with eigenvalue zero. The situation can be intuitively understood using the flow chart shown in Fig.~\ref{fig:setup} (b). As double and triple excitations are prohibited, only four classical states ($\ket{000}$, $\ket{100}$, $\ket{010}$, $\ket{001}$) are relevant in the evolution of the machine. Two of these states, $\ket{000}$ and $\ket{001}$ are directly coupled to reservoir transitions and play no role at long times. The steady state  is pure and a superposition of $\ket{100}$ and $\ket{010}$ \cite{footnote4},
\begin{align}\label{eq:ss}
	\ket{\Psi_{\text{ss}} }= \left(\hspace*{-0.1cm}\frac{g_{23}}{\sqrt{g^2_{23}+g^2_{13}}}\ket{10}-\frac{g_{13}}{\sqrt{g^2_{23}+g^2_{13}}}\ket{01} \hspace*{-0.1cm}\right)\otimes \ket{0},
\end{align}which is actually a 1-particle energy eigenstate
of the Hamiltonian \eqref{eq:ham}. This state can be written as $\ket{\Psi_{\text{ss}}} = \left(\cos\theta\ket{10}-\sin\theta\ket{01} \right)\otimes \ket{0}$, with $\theta\coloneqq\arctan\left( g_{13}/g_{23}\right)$. Clearly, we obtain a partially entangled pure state between qubits 1 and 2, while qubit 3 is pushed into its ground state. The coefficients of the superposed states depend solely on the couplings between the qubits. Setting $g_{13}=g_{23}$,  we obtain a maximally entangled state in the form of the singlet $\ket{\Psi^-}=\left(\ket{10}-\ket{01}\right)/\sqrt{2}$. Importantly, these results are independent of any temperatures of the reservoirs and the coupling rates between the latters and the qubits (within the ideal limit \eqref{eq:lim} and the validity of the master equation).

While Lindbladian evolution typically decoheres an initially pure state into a mixed state, it is generally possible to obtain a pure steady-state provided that certain conditions are satisfied. Firstly, the state $\ket{\Psi_{\text{ss}}}$ is invariant under the action of the jump operators $L_{100}$, $L_{300}$ and $L_{300}^\dagger$ which are relevant for the evolution. In other words, $\ket{\Psi_{\text{ss}}}$ is unaffected by the dissipators in Eq. \eqref{eq:lind} that remain after applying the limits \eqref{eq:lim}. Secondly, this state is an eigenstate of the effective non-Hermitian Hamiltonian $H^\prime=H-i\left( \gamma_{100}^+ L^\dagger_{100}L_{100}+\gamma_{300}^+ L^\dagger_{300}L_{300}+\gamma_{300}^-L_{300}L^\dagger_{300}\right)/2$. These properties here ensure that $\ket{\Psi_{\text{ss}}}$ is the unique steady state of $\mathcal L$ under the considered limits \cite{Yamamoto2005,Kraus2008}.  Using such reasoning, it can be shown  that no two-qubit machine (autonomous or not) in the regime of validity of the local master equation can deterministically generate a pure entangled steady state, implying that our machine is minimal.

\subsection{Minimality of the three-qubit setup }\label{sec:min}

We now argue that a two-qubit machine cannot produce a Bell state as the unique steady state of local Lindbladian evolution considered in past works (e.g.~in Refs.~\cite{Brask2015,Hofer2017,Tacchino2018,Khandelwal2020,Brask2022operational,Prech2023}). We assume a flip-flop interaction between the qubits, and that each qubit is individually coupled to a thermal reservoir. The Hamiltonian of the qubits is given by 
		\begin{align}\label{eq:twoH}
			H=\sum_{i=1,2}\varepsilon\sigma_+^{(i)}\sigma_-^{(i)} + g\left(\sigma_+^{(1)}\sigma_-^{(2)} + \sigma_-^{(1)}\sigma_+^{(2)} \right),
		\end{align}where $\sigma_{\pm}^{(i)}$ are the raisng and lowering operators of qubit $i$. There are a total of eight Lindblad operators corresponding to possible transitions that the reservoirs can induce,
\begin{equation}
	\begin{aligned}\label{eq:Ls}
		&L_{1} = \ketbra{10}{00},\,\, L_2 = \ketbra{01}{00}, \,\, L_3 = \ketbra{01}{11},\,\, L_4 = \ketbra{10}{11}\\
		&L_5 = \ketbra{00}{10},\,\, L_6 = \ketbra{00}{01},\,\, L_7=\ketbra{11}{01},\,\, L_8 = \ketbra{11}{10}.
	\end{aligned}
\end{equation}
We assume that the rates $\gamma_j$ corresponding to these transitions can in general be distinct and can also be zero. The Lindblad equation then takes the form,
\begin{align}\label{eq:lindmin}
	\dot\rho = \mathcal L\rho = -i[H,\rho]+\sum_{j=1}^{8} \gamma_j\mathcal D[L_j]\rho.
\end{align}If there exists a steady state, it must satisfy $\mathcal L\rho_{\text{ss}}=0$. Moreover, in the case of a pure steady state, $\ket{\psi_{\text{ss}}}$, we must have that all the dissipators annihilate this state, i.e., $\mathcal D[L_j]\ketbra{\psi_{\text{ss}}}{\psi_{\text{ss}}}=0$  \cite{Kraus2008}. To satisfy $\mathcal L \ketbra{\psi_{\text{ss}}}{\psi_{\text{ss}}}=0$, we must also have that $-i[H,\ketbra{\psi_{\text{ss}}}{\psi_{\text{ss}}}]=0$, i.e., $\ket{\psi_{\text{ss}}}$ must be an eigenstate of the Hamiltonian. Since the two Bell states $\ket{\Psi^\pm}=(\ket{10}\pm\ket{01})/\sqrt{2}$ are eigenstates of $H$, we  try to see whether these states can be the steady state of Eq. \eqref{eq:lindmin}. First, we note that these Bell states are annihilated by the dissipators $\mathcal D[L_{1-4}]$ but not by $\mathcal D[L_{5-8}]$. Therefore, from Eq. \eqref{eq:lindmin}, we remove $L_{5-8}$ and are left with the following equation to check,
\begin{align}\label{eq:lindmin1}
	\dot\rho = \tilde{\mathcal L}\rho=  -i[H,\rho]+\sum_{j=1}^{4} \gamma_j\mathcal D[L_j]\rho.
\end{align}
We note that while $\tilde{\mathcal L}$ satisfies $\tilde{\mathcal L}\ketbra{\Psi^{\pm}}{\Psi^{\pm}}=0$, it does not have a unique zero eigenvalue and an initial-state-independent steady state in the usual sense. Specifically, in general, $\tilde{\mathcal L}$ has two zero eigenvalues, which  correspond to multiple fixed points of $\tilde{\mathcal L}$. Moreover, due to the structure of the Lindblad operators, there can be residual oscillations even in the steady state (see, for example, Ref. \cite{VVAlbert}). This can be seen with a simple example. Suppose the system starts initially in the tensor product of the qubit ground states, $\ket{00}$. Then the system can gain an excitation on qubit 1 or 2 (through $L_1$ or $L_2$, respectively), but cannot gain a further excitation or lose one. Furthermore, through the flip-flop interaction Hamiltonian, the excitation continuously jumps from qubit 1 and 2, with a frequency that is determined by the coupling strength between the qubits. Therefore, a Bell state cannot be the unique steady state of the two-qubit machine and Eq. \eqref{eq:lindmin}. 

We note that the above discussion can be extended to any two-qubit Hamiltonian. To produce a steady Bell-state, the Hamiltonian must have this state as an eigenstate. The Hamiltonian \eqref{eq:twoH} has $\ket{\Psi^{\pm}}$ as eigenstates. We may instead consider another Hamiltonian that has $\ket{\Phi^\pm}=(\ket{00}\pm\ket{11})/\sqrt{2}$ as eigenstates. It can be checked that dissipators corresponding to the jump operators $L_{5-8}$ in Eq. \eqref{eq:Ls} annihilate $\ket{\Phi^{\pm}}$ into the null state, i.e., $\mathcal D[L_{5-8}]\ketbra{\Phi^\pm}{\Phi^{\pm}}=0$. However, a Lindblad equation with just $L_{5-8}$ leads to similar problems as described above and the evolution is initial state dependent. For example, initial ground $\ket{00}$ and excited $\ket{11}$ states  have no excitation or de-excitation channels, respectively. On the other hand, an initial state like $\ket{10}$ can evolve either to $\ket{00}$ or $\ket{11}$, with no further jump-channel and only the interaction Hamiltonian to control the dynamics. Therefore, no two-qubit Hamiltonian can possibly produce a perfect Bell state as the steady state of the two-qubit thermal machine.

\begin{figure*}
	\centering
	\includegraphics[width=0.85\textwidth]{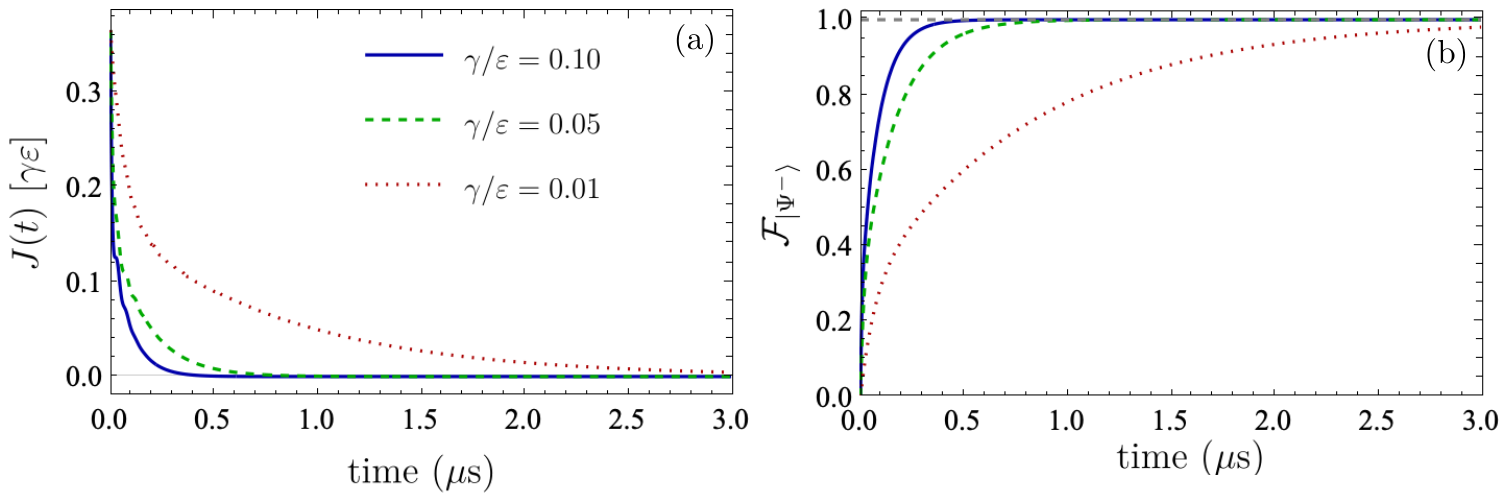}
	\caption{(a) Energy current and (b) fidelity with $\ket{\Psi^-}\otimes \ket{0}$ as a function of time, with the parameters $\varepsilon=1$ GHz, $T_1=T_3=T=\varepsilon$, $\gamma/\varepsilon=\gamma_1/\varepsilon=\gamma_3/\varepsilon$,  $g_{13}/\varepsilon=g_{23}/\varepsilon=0.05$, $U\to\infty$, $\mu\to\infty$, $U/\mu\to\infty$ and system-reservoir couplings as given in the legend.  The qubits are initially in the ground state $\ket{000}$.}
	\label{fig:transient}
	
\end{figure*}

\subsection{Current at ideal operation} 

The average energy current, $J$, can be defined as the average rate of energy exchange between the system and reservoirs. Here, the energy current from reservoir $j$ into the system takes the form \cite{Pottsnotes},{ 
\begin{align}
Q_j(t) \hspace*{-0.1cm}=\hspace*{-0.1cm} \text{Tr}\Bigg\{\hspace*{-0.1cm} H\hspace*{-0.3cm}\sum_{p,q\in\{0,1\}}\hspace*{-0.3cm}\left(\gamma_{jpq}^+ \mathcal D\left[ L_{jpq}\right]\rho\left( t\right)+\gamma_{jpq}^- \mathcal D\left[ L^\dagger_{jpq}\right]\hspace*{-0.1cm}\rho\left( t\right)\right)\hspace*{-0.15cm}\Bigg\}.
\end{align}
In our convention, the heat flow from a reservoir to a qubit is positive, while the other way round is negative. Therefore, currents from the high bias and low bias reservoirs have opposite signs and the average current can be naturally defined as} $J(t)\coloneqq \left(Q_1(t) -Q_3(t)\right)/2$. 

Since at long times, the third qubit is completely depopulated, no excitation can travel into the right reservoir. Moreover, due to the large chemical potential, no excitation can travel back into the left reservoir. As a result, there cannot be any current between the system and reservoirs in the steady state despite the reservoirs being heavily out-of-equilibrium with each other. Therefore, $\ket{\Psi_{\text{ss}}}$ is a dark or non-conducting steady-state of the machine, similar to the works \cite{EmaryPRB2007, WrzesniewskiPRB2018}.

 In Fig. \ref{fig:transient} (a), we plot the energy current as a function of time with the three qubits initially in the ground state $\ket{000}$. This is merely a relevant example, since the steady-state \eqref{eq:ss} holds for any choice of initial state.
As expected, we find that that the initially non-zero energy current drops to zero. We note that a deviation from the ideal limits \eqref{eq:lim} drives the system to a mixed ``bright'' steady state exhibiting a non-zero energy current. However, this state can still be considerably entangled, as discussed below.

We remark that in the absence of dissipation, $\ket{\Psi_{\text{ss}}}$ is a time-independent solution of the Schr\"odinger equation for the three levels $\ket{100}$, $\ket{010}$ and $\ket{001}$. Therefore, these states  effectively form a lambda system with the couplings $g_{13}$ and $g_{23}$ playing the role of Rabi frequencies of a probe field. The three-qubit state is coherently trapped \cite{Lambropoulos2007} between the states $\ket{010}$ and $\ket{100}$ with coefficients determined by the couplings.

\begin{figure}
	\centering
	\includegraphics[width=0.85\columnwidth]{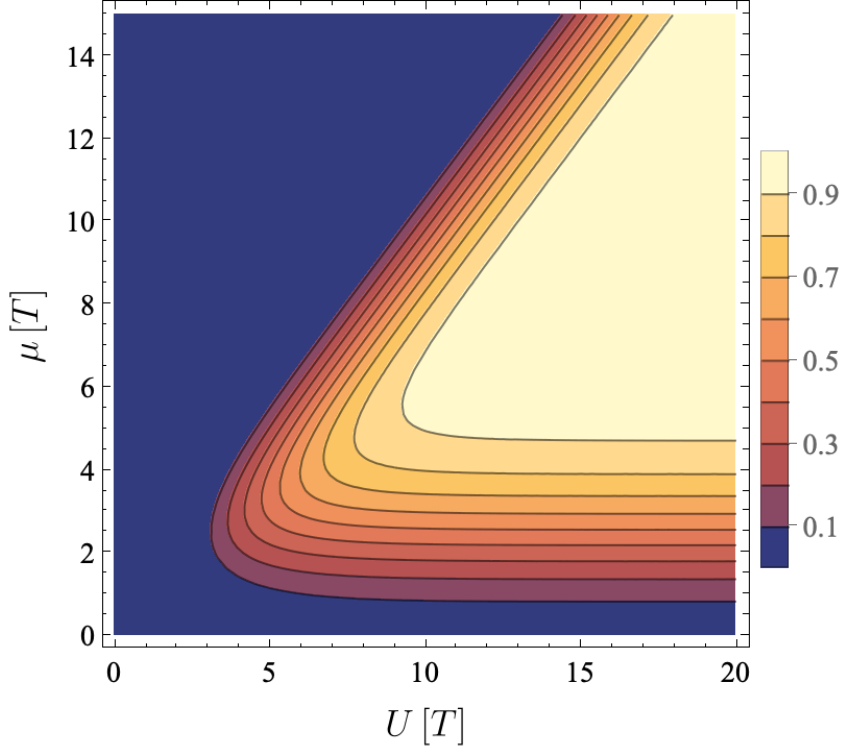}
	\caption{Contour plot of the concurrence between the first two qubits as a function of $\mu$ and $U$, with the system-reservoir couplings $\gamma_1/\varepsilon=\gamma_3/\varepsilon=0.1$ and other parameters taken from Fig. \ref{fig:transient}. Near-maximal entanglement is obtained for a large range of parameters, such that $\mu/T$ and $U/T$ are both sufficiently larger than 1.}
	\label{fig:robust}
\end{figure}

\subsection{Time-scale and entanglement generation away from ideal operation}
 Although the maximally entangled state can be obtained independently of the couplings and temperatures (as long as $g_{13}=g_{23}$), the couplings to the reservoirs determine the time-scale at which the steady state is reached. Specifically, the larger the coupling, the shorter the time-scale. In Fig. \ref{fig:transient} (b), we plot the fidelity of $\rho(t)$ with the state $\ket{\Psi^-}\otimes\ket{0}$, with the system initially in the ground state $\ket{000}$. Choosing the natural frequency of the qubits as $1$ GHz \cite{footnote3}, we find that the steady state can be reached within a few microseconds for all three considered coupling strengths. This is only a relevant example; the relaxation time scale is determined by the Liouvillian eigenvalue having the largest (smallest negative) real part. This is controlled specifically by the system-reservoir couplings. Therefore, this time-scale is the same for any initial state and of the order of magnitude of $1/\gamma$.\par 
In Fig. \ref{fig:robust}, we investigate the quality of the generated entanglement with respect to changes in the system's parameters, i.e.,~away from the limits in Eq.~\eqref{eq:lim}. In such a situation, all possible classical states of the three-qubit system are involved in the dynamics, including at long-times. The solution must therefore take into account all possible transitions induced by the jump operators $L_{jpq}$ and $L_{jpq}^\dagger$ in the general Lindblad equation \eqref{eq:lind}. As a measure of entanglement, we use the concurrence, which for the state of qubits 1 and 2 can be written as $C(\rho)=2\,\text{max}\{0,\lvert c\rvert-\sqrt{p_{11}p_{00}}\}$, where $c$ is the coherence corresponding to element $\ketbra{01}{10}$ and $p_{11}(p_{00})$ is the probability corresponding to the state $\ketbra{11}(\ketbra{00}{00})$. We find that large entanglement can be created with reasonable Coulomb interaction and chemical potentials. As expected from the previous discussion, to generate a large amount of entanglement, we also find that $U$ should be chosen to be sufficiently larger than $\mu$. For instance, choosing $U=15T$ and $\mu=8T$ (where $T_1=T_3=T$) yields a concurrence greater than 0.99, while $\mu=15T$ with the same Coulomb interaction yields only 0.25. This is due to the presence of double excitations coming from the left reservoir when the Coulomb interaction is not large compared with the chemical potential.  Overall, we note that the scheme requires the Coulomb interaction and chemical potential to be much larger than the qubit energies and the temperatures. In the Supplementary Information, we also consider variation in the coupling rates $\gamma_1$ and $\gamma_3$, as well as the influence of single-qubit dephasing.

Our analytical results have been obtained using the Lindblad equation \eqref{eq:lind} with local coupling, which restricts to individual transitions between reservoirs and qubits. In order to assess the validity of this approach, we have shown numerical calculations with the second-order von Neumann approach \cite{Pedersen2005,Pedersen2007} using the QMEQ package \cite{KirsanskasComputPhysCommun2017} in the Supplementary Information. This approach includes cotunneling events, partially lifting the blockade of current in the steady state. For larger system-reservoir coupling, we find only a minor reduction in the entanglement, as quantified by the concurrence (still above 98\% for parameters from Fig.~\ref{fig:robust}). In addition, we find that Lamb-shift terms lead to a slight change in the resonant condition between the qubit energies. Importantly, both effects vanish with decreasing system-reservoir coupling.  More details can be found in the Supplementary Information.

\begin{figure}
	\centering
	\includegraphics[width=1\columnwidth]{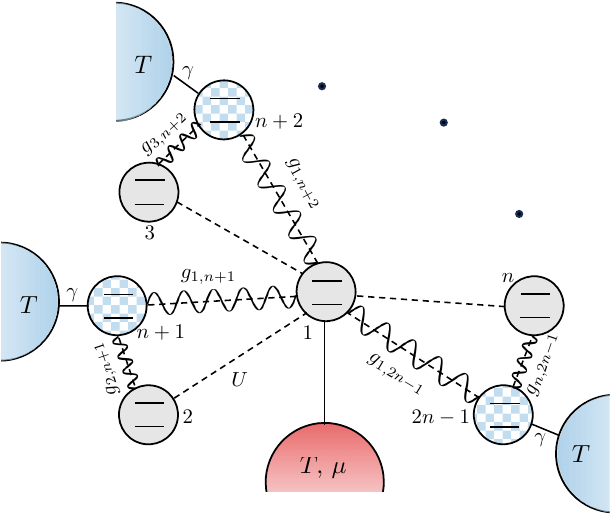}
	\caption{A $(2n-1)$-qubit generalisation of the three-qubit machine.  An $n$-qubit W state is now obtained between qubits $1$,$2$,...,$n$ (gray), while the sink qubits $(n+1)$,$(n+2)$,...,$(2n-1)$ (shaded, blue) are pushed to their ground state. The dashed lines represent Coulomb interaction with strength $U$, and the wiggly lines represent inter-qubit coupling between qubits $i$ and $j$ with strength $g_{i,j}$. For simplicity, we have chosen the temperatures and system-reservoir couplings to be equal. For $n=2$, the machine reduces to the one in Fig.~\ref{fig:setup} (a), producing $\ket{\Psi^-}\otimes\ket{0}$ when the inter-qubit couplings are equal. For each additional qubit to be entangled, one extra sink qubit and one extra ``triangle'' are necessary in the setup.} \label{fig:NGME}
\end{figure}

\subsection{Multipartite entanglement generation} 
A natural question is whether autonomous resources and two-body interactions can go even further and produce maximally entangled states of many qubits. A naive approach is to add a fourth qubit in the system in Fig.~\ref{fig:setup} and coupling it to the sink qubit, {but this does not yield  multipartite entanglement or even a unique steady state. The reason for this is similar to the explanation in Sec. \ref{sec:min}. For successful operation, i.e., to produce a unique pure entangled steady state, the scheme requires coupling to a filled reservoir, as well as an exit way for excitations through the sink qubit. In the absence of such exit ways, there will be oscillations due to inter-qubit couplings even at long times. Therefore, the additional qubit to be entangled requires its own, additional, sink qubit. That is, to generate maximal genuine multipartite entanglement between three qubits, we use two auxiliary qubits serving as sinks. This is illustrated in Fig~\ref{fig:NGME}. This idea can be directly extended to an arbitrary number of qubits; for every additional qubit to be entangled, we couple it to the qubit already connected to the high-bias reservoir, and then we add a corresponding sink qubit. This can be seen as many `triangles' of qubits, within each of which we have an electrostatic interaction between every pair and a flip-flop interaction between the sink qubit and the other two qubits (in complete analogy with the original machine in Fig.~\ref{fig:setup}). In this way, we can deterministically generate an $n$-qubit W-type entangled state in a $(2n-1)$-qubit autonomous thermal machine.  Under the same limits as the two-qubit case, namely Eq.~\eqref{eq:lim}, we show in the Supplementary Information that this scheme corresponds to a generalised coherent population trapping over $n$ states (instead of just two), which renders the steady state of this scheme pure and entangled. Specifically, it is
		\begin{align}\label{eq:genss}
			\ket{\Psi^n_\text{ss}} = \frac{1}{\sqrt{\beta}}\hspace*{-0.1cm}\left(\hspace*{-0.1cm} \ket{10...0}-\hspace*{-0.1cm}\sum_{j=1}^{n-1} \alpha_j \ket{\bar0_j}\hspace*{-0.1cm}\otimes\hspace*{-0.1cm}\ket{1}\hspace*{-0.1cm}\otimes\hspace*{-0.1cm}\ket{\bar 0_{n-j-1}}\hspace*{-0.1cm} \right)\hspace*{-0.1cm}\otimes\ket{\bar 0_{n-1}}\hspace*{-0.1cm},
		\end{align}
		where $\ket{\bar 0_{k}}\coloneqq \ket{0}\otimes\ket{0}\otimes\cdot\cdot\cdot\otimes\ket{0}$ is the ground state of $k$ qubits and
	\begin{align}
	\alpha_j \coloneqq \frac{g_{1,n+j}}{g_{1+j,n+j}}\quad\text{and}\quad \beta \coloneqq1+\sum_{j=1}^{n-1}\alpha_j^2.
\end{align}$\ket{\Psi^n_{\text{ss}}}$ corresponds to a W-type partially entangled state for qubits $1-n$. The condition to obtain maximal entanglement is similar to the three-qubit machine. Here, if the inter-qubit couplings within each triangle are equal, i.e., $g_{1,n+j}=g_{1+j,n+j}$ or $\alpha_j=1$, $\ket{\Psi^n_{\text{ss}}}$ corresponds exactly to the following W state of $n$ qubits, while $n-1$ qubits are pushed into their ground state,
		
\begin{equation}
\ket{W_n}=\hspace*{-0.1cm}\frac{1}{\hspace*{-0.03cm}\sqrt{\hspace*{-0.05cm}n}}\hspace*{-0.1cm}\left(\ket{100... 0}-\ket{010... 0 }-...-\ket{000... 1}\hspace*{-0.05cm}\right)\otimes \ket{\bar{0}_{n-1}}
\end{equation} $\ket{W_n}$ denotes a W state in the space of $n$ qubits and a ground state in the next $n-1$ qubits. In this enumeration, the first system is the high-bias qubit and the ground state $\ket{\bar{0}_{n-1}}$ corresponds to all the sink qubits.  For simplicity, in Fig.~\ref{fig:NGME}, we have chosen all temperatures and all qubit-reservoir couplings to be equal. However, this is not a necessary condition to produce a $\ket{W_n}$ state. The couplings can be chosen almost arbitrarily (within the validity of the Lindblad equation) and the temperatures have to be chosen such  that $U\gg \mu\gg T_j$. Robustness to variations in system-reservoir couplings are further discussed in the Supplementary Information.

\section{Discussion}

A considerable number of earlier works on autonomous entanglement generation focussed on creating two-qubit entanglement using a setup of two qubits. The amount of entanglement in these works was always noisy and far from maximal. In this work, we have shown that this is limited due to the structure of the Lindblad equation - it is impossible to generate a perfect Bell state using a a two-qubit autonomous thermal machine. Importantly, we have introduced a minimal three-qubit architecture that generates a steady Bell state for ideal operation. The scheme is robust; even away from ideal operation, it can generate near-maximal entanglement. Furthermore, we have demonstrated that our results can be extended to producing genuinely multipartite entanglement in the form of W states of an arbitrary number of qubits.  It is an interesting theoretical question whether our ideas can be extended to produce arbitrary pure entanglement, in particular the Greenberger-Horne-Zeilinger states \cite{Dur2000} and whether it is possible to obtain high-dimensional entanglement \cite{Cobucci2024} in a similar setting.

While maximal entanglement generation is possible with non-autonomous resources such as driving \cite{Khandelwal2023} and athermality \cite{Oliveira2024}, our work provides a fully autonomous pathway to generating maximal entanglement. In other words, our work demonstrates that pure dissipation into thermal environments is sufficient to generate the strongest forms of quantum correlations. This reveals the striking fundamental power of autonomous evolution.

Finally, beyond their fundamental significance, we believe that recent developments in quantum technologies make our predictions experimentally feasible. While there are many platforms that may be suitable, electronic quantum dots are a natural candidate. Here, the right coupling regimes (within the validity of our master equation) can be already engineered in triple-dot setups \cite{GaudreauPRL2006,RoggeNJP2009}. Crucially, Coulomb repulsion is naturally present and bias voltage (and therefore the chemical potential) can freely be controlled, making the operation of the setup possible close to the ideal limit.

\section*{Code availability}
The code used for the 2vN approach is available at \href{https://github.com/gedaskir/qmeq}{https://github.com/gedaskir/qmeq}.

\begin{acknowledgments}
We thank Peter Samuelsson, Claudio Verdozzi and Patrick Potts for discussions and feedback. This work is supported by the Wenner-Gren Foundation and by the Knut and Alice Wallenberg Foundation through the Wallenberg Center for Quantum Technology (WACQT). SK further acknowledges the Swiss National Science Foundation Grant No. P500PT\textunderscore222265. BAA was supported by the Swedish Research Council, Grant No. 2018-03921.
\end{acknowledgments}



\begin{thebibliography}{64}%
	\makeatletter
	\providecommand \@ifxundefined [1]{%
		\@ifx{#1\undefined}
	}%
	\providecommand \@ifnum [1]{%
		\ifnum #1\expandafter \@firstoftwo
		\else \expandafter \@secondoftwo
		\fi
	}%
	\providecommand \@ifx [1]{%
		\ifx #1\expandafter \@firstoftwo
		\else \expandafter \@secondoftwo
		\fi
	}%
	\providecommand \natexlab [1]{#1}%
	\providecommand \enquote  [1]{``#1''}%
	\providecommand \bibnamefont  [1]{#1}%
	\providecommand \bibfnamefont [1]{#1}%
	\providecommand \citenamefont [1]{#1}%
	\providecommand \href@noop [0]{\@secondoftwo}%
	\providecommand \href [0]{\begingroup \@sanitize@url \@href}%
	\providecommand \@href[1]{\@@startlink{#1}\@@href}%
	\providecommand \@@href[1]{\endgroup#1\@@endlink}%
	\providecommand \@sanitize@url [0]{\catcode `\\12\catcode `\$12\catcode
		`\&12\catcode `\#12\catcode `\^12\catcode `\_12\catcode `\%12\relax}%
	\providecommand \@@startlink[1]{}%
	\providecommand \@@endlink[0]{}%
	\providecommand \url  [0]{\begingroup\@sanitize@url \@url }%
	\providecommand \@url [1]{\endgroup\@href {#1}{\urlprefix }}%
	\providecommand \urlprefix  [0]{URL }%
	\providecommand \Eprint [0]{\href }%
	\providecommand \doibase [0]{https://doi.org/}%
	\providecommand \selectlanguage [0]{\@gobble}%
	\providecommand \bibinfo  [0]{\@secondoftwo}%
	\providecommand \bibfield  [0]{\@secondoftwo}%
	\providecommand \translation [1]{[#1]}%
	\providecommand \BibitemOpen [0]{}%
	\providecommand \bibitemStop [0]{}%
	\providecommand \bibitemNoStop [0]{.\EOS\space}%
	\providecommand \EOS [0]{\spacefactor3000\relax}%
	\providecommand \BibitemShut  [1]{\csname bibitem#1\endcsname}%
	\let\auto@bib@innerbib\@empty
	\bibitem [{\citenamefont {Scovil}\ and\ \citenamefont
		{Schulz-DuBois}(1959)}]{Scovil1959}%
	\BibitemOpen
	\bibfield  {author} {\bibinfo {author} {\bibfnamefont {H.~E.~D.}\
			\bibnamefont {Scovil}}\ and\ \bibinfo {author} {\bibfnamefont {E.~O.}\
			\bibnamefont {Schulz-DuBois}},\ }\bibfield  {title} {\bibinfo {title}
		{Three-level masers as heat engines},\ }\href
	{https://doi.org/10.1103/PhysRevLett.2.262} {\bibfield  {journal} {\bibinfo
			{journal} {Phys. Rev. Lett.}\ }\textbf {\bibinfo {volume} {2}},\ \bibinfo
		{pages} {262} (\bibinfo {year} {1959})}\BibitemShut {NoStop}%
	\bibitem [{\citenamefont {Alicki}(1979)}]{Alicki1979}%
	\BibitemOpen
	\bibfield  {author} {\bibinfo {author} {\bibfnamefont {R.}~\bibnamefont
			{Alicki}},\ }\bibfield  {title} {\bibinfo {title} {The quantum open system as
			a model of the heat engine},\ }\href
	{https://dx.doi.org/10.1088/0305-4470/12/5/007} {\bibfield  {journal}
		{\bibinfo  {journal} {J. Phys A: Math. Gen.}\ }\textbf {\bibinfo {volume}
			{12}},\ \bibinfo {pages} {L103} (\bibinfo {year} {1979})}\BibitemShut
	{NoStop}%
	\bibitem [{\citenamefont {Kosloff}\ and\ \citenamefont
		{Levy}(2014)}]{Kosloff2014}%
	\BibitemOpen
	\bibfield  {author} {\bibinfo {author} {\bibfnamefont {R.}~\bibnamefont
			{Kosloff}}\ and\ \bibinfo {author} {\bibfnamefont {A.}~\bibnamefont {Levy}},\
	}\bibfield  {title} {\bibinfo {title} {Quantum heat engines and
			refrigerators: Continuous devices},\ }\href
	{https://doi.org/10.1146/annurev-physchem-040513-103724} {\bibfield
		{journal} {\bibinfo  {journal} {Annu. Rev. Phys. Chem.}\ }\textbf {\bibinfo
			{volume} {65}},\ \bibinfo {pages} {365} (\bibinfo {year} {2014})}\BibitemShut
	{NoStop}%
	\bibitem [{\citenamefont {Mitchison}(2019)}]{Mitchison2019}%
	\BibitemOpen
	\bibfield  {author} {\bibinfo {author} {\bibfnamefont {M.~T.}\ \bibnamefont
			{Mitchison}},\ }\bibfield  {title} {\bibinfo {title} {Quantum thermal
			absorption machines: refrigertors, engines and clocks},\ }\href
	{https://doi.org/10.1080/00107514.2019.1631555} {\bibfield  {journal}
		{\bibinfo  {journal} {Contemp. Phys.}\ }\textbf {\bibinfo {volume} {60}},\
		\bibinfo {pages} {164} (\bibinfo {year} {2019})}\BibitemShut {NoStop}%
	\bibitem [{\citenamefont {Bhattacharjee}\ and\ \citenamefont
		{Dutta}(2021)}]{Bhattacharjee2021}%
	\BibitemOpen
	\bibfield  {author} {\bibinfo {author} {\bibfnamefont {S.}~\bibnamefont
			{Bhattacharjee}}\ and\ \bibinfo {author} {\bibfnamefont {A.}~\bibnamefont
			{Dutta}},\ }\bibfield  {title} {\bibinfo {title} {Quantum thermal machines
			and batteries},\ }\href {https://doi.org/10.1140/epjb/s10051-021-00235-3}
	{\bibfield  {journal} {\bibinfo  {journal} {Eur. Phys. J. B}\ }\textbf
		{\bibinfo {volume} {94}},\ \bibinfo {pages} {239} (\bibinfo {year}
		{2021})}\BibitemShut {NoStop}%
	\bibitem [{\citenamefont {Ro\ss{}nagel}\ \emph {et~al.}(2014)\citenamefont
		{Ro\ss{}nagel}, \citenamefont {Abah}, \citenamefont {Schmidt-Kaler},
		\citenamefont {Singer},\ and\ \citenamefont {Lutz}}]{Rosnagel2014}%
	\BibitemOpen
	\bibfield  {author} {\bibinfo {author} {\bibfnamefont {J.}~\bibnamefont
			{Ro\ss{}nagel}}, \bibinfo {author} {\bibfnamefont {O.}~\bibnamefont {Abah}},
		\bibinfo {author} {\bibfnamefont {F.}~\bibnamefont {Schmidt-Kaler}}, \bibinfo
		{author} {\bibfnamefont {K.}~\bibnamefont {Singer}},\ and\ \bibinfo {author}
		{\bibfnamefont {E.}~\bibnamefont {Lutz}},\ }\bibfield  {title} {\bibinfo
		{title} {Nanoscale heat engine beyond the carnot limit},\ }\href
	{https://doi.org/10.1103/PhysRevLett.112.030602} {\bibfield  {journal}
		{\bibinfo  {journal} {Phys. Rev. Lett.}\ }\textbf {\bibinfo {volume} {112}},\
		\bibinfo {pages} {030602} (\bibinfo {year} {2014})}\BibitemShut {NoStop}%
	\bibitem [{\citenamefont {Roßnagel}\ \emph {et~al.}(2016)\citenamefont
		{Roßnagel}, \citenamefont {Dawkins}, \citenamefont {Tolazzi}, \citenamefont
		{Abah}, \citenamefont {Lutz}, \citenamefont {Schmidt-Kaler},\ and\
		\citenamefont {Singer}}]{Rosnagel2016}%
	\BibitemOpen
	\bibfield  {author} {\bibinfo {author} {\bibfnamefont {J.}~\bibnamefont
			{Roßnagel}}, \bibinfo {author} {\bibfnamefont {S.~T.}\ \bibnamefont
			{Dawkins}}, \bibinfo {author} {\bibfnamefont {K.~N.}\ \bibnamefont
			{Tolazzi}}, \bibinfo {author} {\bibfnamefont {O.}~\bibnamefont {Abah}},
		\bibinfo {author} {\bibfnamefont {E.}~\bibnamefont {Lutz}}, \bibinfo {author}
		{\bibfnamefont {F.}~\bibnamefont {Schmidt-Kaler}},\ and\ \bibinfo {author}
		{\bibfnamefont {K.}~\bibnamefont {Singer}},\ }\bibfield  {title} {\bibinfo
		{title} {A single-atom heat engine},\ }\href
	{https://doi.org/10.1126/science.aad6320} {\bibfield  {journal} {\bibinfo
			{journal} {Science}\ }\textbf {\bibinfo {volume} {352}},\ \bibinfo {pages}
		{325} (\bibinfo {year} {2016})}\BibitemShut {NoStop}%
	\bibitem [{\citenamefont {Maslennikov}\ \emph {et~al.}(2019)\citenamefont
		{Maslennikov}, \citenamefont {Ding}, \citenamefont {Habl{\"u}tzel},
		\citenamefont {Gan}, \citenamefont {Roulet}, \citenamefont {Nimmrichter},
		\citenamefont {Dai}, \citenamefont {Scarani},\ and\ \citenamefont
		{Matsukevich}}]{Maslennikov2019}%
	\BibitemOpen
	\bibfield  {author} {\bibinfo {author} {\bibfnamefont {G.}~\bibnamefont
			{Maslennikov}}, \bibinfo {author} {\bibfnamefont {S.}~\bibnamefont {Ding}},
		\bibinfo {author} {\bibfnamefont {R.}~\bibnamefont {Habl{\"u}tzel}}, \bibinfo
		{author} {\bibfnamefont {J.}~\bibnamefont {Gan}}, \bibinfo {author}
		{\bibfnamefont {A.}~\bibnamefont {Roulet}}, \bibinfo {author} {\bibfnamefont
			{S.}~\bibnamefont {Nimmrichter}}, \bibinfo {author} {\bibfnamefont
			{J.}~\bibnamefont {Dai}}, \bibinfo {author} {\bibfnamefont {V.}~\bibnamefont
			{Scarani}},\ and\ \bibinfo {author} {\bibfnamefont {D.}~\bibnamefont
			{Matsukevich}},\ }\bibfield  {title} {\bibinfo {title} {Quantum absorption
			refrigerator with trapped ions},\ }\href
	{https://doi.org/10.1038/s41467-018-08090-0} {\bibfield  {journal} {\bibinfo
			{journal} {Nat. Commun.}\ }\textbf {\bibinfo {volume} {10}},\ \bibinfo
		{pages} {202} (\bibinfo {year} {2019})}\BibitemShut {NoStop}%
	\bibitem [{\citenamefont {Klatzow}\ \emph {et~al.}(2019)\citenamefont
		{Klatzow}, \citenamefont {Becker}, \citenamefont {Ledingham}, \citenamefont
		{Weinzetl}, \citenamefont {Kaczmarek}, \citenamefont {Saunders},
		\citenamefont {Nunn}, \citenamefont {Walmsley}, \citenamefont {Uzdin},\ and\
		\citenamefont {Poem}}]{Klatzow2019}%
	\BibitemOpen
	\bibfield  {author} {\bibinfo {author} {\bibfnamefont {J.}~\bibnamefont
			{Klatzow}}, \bibinfo {author} {\bibfnamefont {J.~N.}\ \bibnamefont {Becker}},
		\bibinfo {author} {\bibfnamefont {P.~M.}\ \bibnamefont {Ledingham}}, \bibinfo
		{author} {\bibfnamefont {C.}~\bibnamefont {Weinzetl}}, \bibinfo {author}
		{\bibfnamefont {K.~T.}\ \bibnamefont {Kaczmarek}}, \bibinfo {author}
		{\bibfnamefont {D.~J.}\ \bibnamefont {Saunders}}, \bibinfo {author}
		{\bibfnamefont {J.}~\bibnamefont {Nunn}}, \bibinfo {author} {\bibfnamefont
			{I.~A.}\ \bibnamefont {Walmsley}}, \bibinfo {author} {\bibfnamefont
			{R.}~\bibnamefont {Uzdin}},\ and\ \bibinfo {author} {\bibfnamefont
			{E.}~\bibnamefont {Poem}},\ }\bibfield  {title} {\bibinfo {title}
		{Experimental demonstration of quantum effects in the operation of
			microscopic heat engines},\ }\href
	{https://doi.org/10.1103/PhysRevLett.122.110601} {\bibfield  {journal}
		{\bibinfo  {journal} {Phys. Rev. Lett.}\ }\textbf {\bibinfo {volume} {122}},\
		\bibinfo {pages} {110601} (\bibinfo {year} {2019})}\BibitemShut {NoStop}%
	\bibitem [{\citenamefont {Pearson}\ \emph {et~al.}(2021)\citenamefont
		{Pearson}, \citenamefont {Guryanova}, \citenamefont {Erker}, \citenamefont
		{Laird}, \citenamefont {Briggs}, \citenamefont {Huber},\ and\ \citenamefont
		{Ares}}]{Pearson2021}%
	\BibitemOpen
	\bibfield  {author} {\bibinfo {author} {\bibfnamefont {A.~N.}\ \bibnamefont
			{Pearson}}, \bibinfo {author} {\bibfnamefont {Y.}~\bibnamefont {Guryanova}},
		\bibinfo {author} {\bibfnamefont {P.}~\bibnamefont {Erker}}, \bibinfo
		{author} {\bibfnamefont {E.~A.}\ \bibnamefont {Laird}}, \bibinfo {author}
		{\bibfnamefont {G.~A.~D.}\ \bibnamefont {Briggs}}, \bibinfo {author}
		{\bibfnamefont {M.}~\bibnamefont {Huber}},\ and\ \bibinfo {author}
		{\bibfnamefont {N.}~\bibnamefont {Ares}},\ }\bibfield  {title} {\bibinfo
		{title} {Measuring the thermodynamic cost of timekeeping},\ }\href
	{https://doi.org/10.1103/PhysRevX.11.021029} {\bibfield  {journal} {\bibinfo
			{journal} {Phys. Rev. X}\ }\textbf {\bibinfo {volume} {11}},\ \bibinfo
		{pages} {021029} (\bibinfo {year} {2021})}\BibitemShut {NoStop}%
	\bibitem [{\citenamefont {Diehl}\ \emph {et~al.}(2008)\citenamefont {Diehl},
		\citenamefont {Micheli}, \citenamefont {Kantian}, \citenamefont {Kraus},
		\citenamefont {B{\"u}chler},\ and\ \citenamefont {Zoller}}]{Diehl2008}%
	\BibitemOpen
	\bibfield  {author} {\bibinfo {author} {\bibfnamefont {S.}~\bibnamefont
			{Diehl}}, \bibinfo {author} {\bibfnamefont {A.}~\bibnamefont {Micheli}},
		\bibinfo {author} {\bibfnamefont {A.}~\bibnamefont {Kantian}}, \bibinfo
		{author} {\bibfnamefont {B.}~\bibnamefont {Kraus}}, \bibinfo {author}
		{\bibfnamefont {H.~P.}\ \bibnamefont {B{\"u}chler}},\ and\ \bibinfo {author}
		{\bibfnamefont {P.}~\bibnamefont {Zoller}},\ }\bibfield  {title} {\bibinfo
		{title} {Quantum states and phases in driven open quantum systems with cold
			atoms},\ }\href {https://doi.org/10.1038/nphys1073} {\bibfield  {journal}
		{\bibinfo  {journal} {Nat. Phys.}\ }\textbf {\bibinfo {volume} {4}},\
		\bibinfo {pages} {878} (\bibinfo {year} {2008})}\BibitemShut {NoStop}%
	\bibitem [{\citenamefont {Verstraete}\ \emph {et~al.}(2009)\citenamefont
		{Verstraete}, \citenamefont {Wolf},\ and\ \citenamefont
		{Ignacio~Cirac}}]{Verstraete2009}%
	\BibitemOpen
	\bibfield  {author} {\bibinfo {author} {\bibfnamefont {F.}~\bibnamefont
			{Verstraete}}, \bibinfo {author} {\bibfnamefont {M.~M.}\ \bibnamefont
			{Wolf}},\ and\ \bibinfo {author} {\bibfnamefont {J.}~\bibnamefont
			{Ignacio~Cirac}},\ }\bibfield  {title} {\bibinfo {title} {Quantum computation
			and quantum-state engineering driven by dissipation},\ }\href
	{https://doi.org/10.1038/nphys1342} {\bibfield  {journal} {\bibinfo
			{journal} {Nat. Phys.}\ }\textbf {\bibinfo {volume} {5}},\ \bibinfo {pages}
		{633} (\bibinfo {year} {2009})}\BibitemShut {NoStop}%
	\bibitem [{\citenamefont {Kastoryano}\ \emph {et~al.}(2011)\citenamefont
		{Kastoryano}, \citenamefont {Reiter},\ and\ \citenamefont
		{S\o{}rensen}}]{Kastoryano2011}%
	\BibitemOpen
	\bibfield  {author} {\bibinfo {author} {\bibfnamefont {M.~J.}\ \bibnamefont
			{Kastoryano}}, \bibinfo {author} {\bibfnamefont {F.}~\bibnamefont {Reiter}},\
		and\ \bibinfo {author} {\bibfnamefont {A.~S.}\ \bibnamefont {S\o{}rensen}},\
	}\bibfield  {title} {\bibinfo {title} {Dissipative preparation of
			entanglement in optical cavities},\ }\href
	{https://doi.org/10.1103/PhysRevLett.106.090502} {\bibfield  {journal}
		{\bibinfo  {journal} {Phys. Rev. Lett.}\ }\textbf {\bibinfo {volume} {106}},\
		\bibinfo {pages} {090502} (\bibinfo {year} {2011})}\BibitemShut {NoStop}%
	\bibitem [{\citenamefont {Krauter}\ \emph {et~al.}(2011)\citenamefont
		{Krauter}, \citenamefont {Muschik}, \citenamefont {Jensen}, \citenamefont
		{Wasilewski}, \citenamefont {Petersen}, \citenamefont {Cirac},\ and\
		\citenamefont {Polzik}}]{Krauter2011}%
	\BibitemOpen
	\bibfield  {author} {\bibinfo {author} {\bibfnamefont {H.}~\bibnamefont
			{Krauter}}, \bibinfo {author} {\bibfnamefont {C.~A.}\ \bibnamefont
			{Muschik}}, \bibinfo {author} {\bibfnamefont {K.}~\bibnamefont {Jensen}},
		\bibinfo {author} {\bibfnamefont {W.}~\bibnamefont {Wasilewski}}, \bibinfo
		{author} {\bibfnamefont {J.~M.}\ \bibnamefont {Petersen}}, \bibinfo {author}
		{\bibfnamefont {J.~I.}\ \bibnamefont {Cirac}},\ and\ \bibinfo {author}
		{\bibfnamefont {E.~S.}\ \bibnamefont {Polzik}},\ }\bibfield  {title}
	{\bibinfo {title} {Entanglement generated by dissipation and steady state
			entanglement of two macroscopic objects},\ }\href
	{https://doi.org/10.1103/PhysRevLett.107.080503} {\bibfield  {journal}
		{\bibinfo  {journal} {Phys. Rev. Lett.}\ }\textbf {\bibinfo {volume} {107}},\
		\bibinfo {pages} {080503} (\bibinfo {year} {2011})}\BibitemShut {NoStop}%
	\bibitem [{\citenamefont {Lin}\ \emph {et~al.}(2013)\citenamefont {Lin},
		\citenamefont {Gaebler}, \citenamefont {Reiter}, \citenamefont {Tan},
		\citenamefont {Bowler}, \citenamefont {S{\o}rensen}, \citenamefont
		{Leibfried},\ and\ \citenamefont {Wineland}}]{Lin2013}%
	\BibitemOpen
	\bibfield  {author} {\bibinfo {author} {\bibfnamefont {Y.}~\bibnamefont
			{Lin}}, \bibinfo {author} {\bibfnamefont {J.~P.}\ \bibnamefont {Gaebler}},
		\bibinfo {author} {\bibfnamefont {F.}~\bibnamefont {Reiter}}, \bibinfo
		{author} {\bibfnamefont {T.~R.}\ \bibnamefont {Tan}}, \bibinfo {author}
		{\bibfnamefont {R.}~\bibnamefont {Bowler}}, \bibinfo {author} {\bibfnamefont
			{A.~S.}\ \bibnamefont {S{\o}rensen}}, \bibinfo {author} {\bibfnamefont
			{D.}~\bibnamefont {Leibfried}},\ and\ \bibinfo {author} {\bibfnamefont
			{D.~J.}\ \bibnamefont {Wineland}},\ }\bibfield  {title} {\bibinfo {title}
		{Dissipative production of a maximally entangled steady state of two quantum
			bits},\ }\href {https://doi.org/10.1038/nature12801} {\bibfield  {journal}
		{\bibinfo  {journal} {Nature}\ }\textbf {\bibinfo {volume} {504}},\ \bibinfo
		{pages} {415} (\bibinfo {year} {2013})}\BibitemShut {NoStop}%
	\bibitem [{\citenamefont {Shankar}\ \emph {et~al.}(2013)\citenamefont
		{Shankar}, \citenamefont {Hatridge}, \citenamefont {Leghtas}, \citenamefont
		{Sliwa}, \citenamefont {Narla}, \citenamefont {Vool}, \citenamefont {Girvin},
		\citenamefont {Frunzio}, \citenamefont {Mirrahimi},\ and\ \citenamefont
		{Devoret}}]{Shankar2013}%
	\BibitemOpen
	\bibfield  {author} {\bibinfo {author} {\bibfnamefont {S.}~\bibnamefont
			{Shankar}}, \bibinfo {author} {\bibfnamefont {M.}~\bibnamefont {Hatridge}},
		\bibinfo {author} {\bibfnamefont {Z.}~\bibnamefont {Leghtas}}, \bibinfo
		{author} {\bibfnamefont {K.~M.}\ \bibnamefont {Sliwa}}, \bibinfo {author}
		{\bibfnamefont {A.}~\bibnamefont {Narla}}, \bibinfo {author} {\bibfnamefont
			{U.}~\bibnamefont {Vool}}, \bibinfo {author} {\bibfnamefont {S.~M.}\
			\bibnamefont {Girvin}}, \bibinfo {author} {\bibfnamefont {L.}~\bibnamefont
			{Frunzio}}, \bibinfo {author} {\bibfnamefont {M.}~\bibnamefont {Mirrahimi}},\
		and\ \bibinfo {author} {\bibfnamefont {M.~H.}\ \bibnamefont {Devoret}},\
	}\bibfield  {title} {\bibinfo {title} {Autonomously stabilized entanglement
			between two superconducting quantum bits},\ }\href
	{https://doi.org/10.1038/nature12802} {\bibfield  {journal} {\bibinfo
			{journal} {Nature}\ }\textbf {\bibinfo {volume} {504}},\ \bibinfo {pages}
		{419} (\bibinfo {year} {2013})}\BibitemShut {NoStop}%
	\bibitem [{\citenamefont {Mancilla}\ \emph {et~al.}(2009)\citenamefont
		{Mancilla}, \citenamefont {Rey-González},\ and\ \citenamefont
		{Fonseca-Romero}}]{Mancilla2009}%
	\BibitemOpen
	\bibfield  {author} {\bibinfo {author} {\bibfnamefont {R.~D.~G.}\
			\bibnamefont {Mancilla}}, \bibinfo {author} {\bibfnamefont {R.~R.}\
			\bibnamefont {Rey-González}},\ and\ \bibinfo {author} {\bibfnamefont
			{K.~M.}\ \bibnamefont {Fonseca-Romero}},\ }\href
	{https://doi.org/10.1088/1751-8113/42/10/105302} {\bibfield  {journal}
		{\bibinfo  {journal} {J. Phys. A: Math. Theor.}\ }\textbf {\bibinfo {volume}
			{42}},\ \bibinfo {pages} {105302} (\bibinfo {year} {2009})}\BibitemShut
	{NoStop}%
	\bibitem [{\citenamefont {Rao}\ and\ \citenamefont
		{M\o{}lmer}(2013)}]{Rao2013}%
	\BibitemOpen
	\bibfield  {author} {\bibinfo {author} {\bibfnamefont {D.~D.~B.}\
			\bibnamefont {Rao}}\ and\ \bibinfo {author} {\bibfnamefont {K.}~\bibnamefont
			{M\o{}lmer}},\ }\bibfield  {title} {\bibinfo {title} {Dark entangled steady
			states of interacting rydberg atoms},\ }\href
	{https://doi.org/10.1103/PhysRevLett.111.033606} {\bibfield  {journal}
		{\bibinfo  {journal} {Phys. Rev. Lett.}\ }\textbf {\bibinfo {volume} {111}},\
		\bibinfo {pages} {033606} (\bibinfo {year} {2013})}\BibitemShut {NoStop}%
	\bibitem [{\citenamefont {Khandelwal}\ \emph {et~al.}(2024)\citenamefont
		{Khandelwal}, \citenamefont {Chen}, \citenamefont {Murch},\ and\
		\citenamefont {Haack}}]{Khandelwal2023}%
	\BibitemOpen
	\bibfield  {author} {\bibinfo {author} {\bibfnamefont {S.}~\bibnamefont
			{Khandelwal}}, \bibinfo {author} {\bibfnamefont {W.}~\bibnamefont {Chen}},
		\bibinfo {author} {\bibfnamefont {K.~W.}\ \bibnamefont {Murch}},\ and\
		\bibinfo {author} {\bibfnamefont {G.}~\bibnamefont {Haack}},\ }\bibfield
	{title} {\bibinfo {title} {Chiral bell-state transfer via dissipative
			liouvillian dynamics},\ }\href
	{https://doi.org/10.1103/PhysRevLett.133.070403} {\bibfield  {journal}
		{\bibinfo  {journal} {Phys. Rev. Lett.}\ }\textbf {\bibinfo {volume} {133}},\
		\bibinfo {pages} {070403} (\bibinfo {year} {2024})}\BibitemShut {NoStop}%
	\bibitem [{\citenamefont {Banerjee}\ \emph {et~al.}(2010)\citenamefont
		{Banerjee}, \citenamefont {Ravishankar},\ and\ \citenamefont
		{Srikanth}}]{Banerjee2010}%
	\BibitemOpen
	\bibfield  {author} {\bibinfo {author} {\bibfnamefont {S.}~\bibnamefont
			{Banerjee}}, \bibinfo {author} {\bibfnamefont {V.}~\bibnamefont
			{Ravishankar}},\ and\ \bibinfo {author} {\bibfnamefont {R.}~\bibnamefont
			{Srikanth}},\ }\bibfield  {title} {\bibinfo {title} {Dynamics of entanglement
			in two-qubit open system interacting with a squeezed thermal bath via
			dissipative interaction},\ }\href
	{https://doi.org/https://doi.org/10.1016/j.aop.2010.01.003} {\bibfield
		{journal} {\bibinfo  {journal} {Ann. Phys.}\ }\textbf {\bibinfo {volume}
			{325}},\ \bibinfo {pages} {816} (\bibinfo {year} {2010})}\BibitemShut
	{NoStop}%
	\bibitem [{\citenamefont {Reiter}\ \emph {et~al.}(2012)\citenamefont {Reiter},
		\citenamefont {Kastoryano},\ and\ \citenamefont {Sørensen}}]{Reiter2012}%
	\BibitemOpen
	\bibfield  {author} {\bibinfo {author} {\bibfnamefont {F.}~\bibnamefont
			{Reiter}}, \bibinfo {author} {\bibfnamefont {M.~J.}\ \bibnamefont
			{Kastoryano}},\ and\ \bibinfo {author} {\bibfnamefont {A.~S.}\ \bibnamefont
			{Sørensen}},\ }\bibfield  {title} {\bibinfo {title} {Driving two atoms in an
			optical cavity into an entangled steady state using engineered decay},\
	}\href {https://doi.org/10.1088/1367-2630/14/5/053022} {\bibfield  {journal}
		{\bibinfo  {journal} {New J. Phys.}\ }\textbf {\bibinfo {volume} {14}},\
		\bibinfo {pages} {053022} (\bibinfo {year} {2012})}\BibitemShut {NoStop}%
	\bibitem [{\citenamefont {Tacchino}\ \emph {et~al.}(2018)\citenamefont
		{Tacchino}, \citenamefont {Auff\`eves}, \citenamefont {Santos},\ and\
		\citenamefont {Gerace}}]{Tacchino2018}%
	\BibitemOpen
	\bibfield  {author} {\bibinfo {author} {\bibfnamefont {F.}~\bibnamefont
			{Tacchino}}, \bibinfo {author} {\bibfnamefont {A.}~\bibnamefont
			{Auff\`eves}}, \bibinfo {author} {\bibfnamefont {M.~F.}\ \bibnamefont
			{Santos}},\ and\ \bibinfo {author} {\bibfnamefont {D.}~\bibnamefont
			{Gerace}},\ }\bibfield  {title} {\bibinfo {title} {Steady state entanglement
			beyond thermal limits},\ }\href
	{https://doi.org/10.1103/PhysRevLett.120.063604} {\bibfield  {journal}
		{\bibinfo  {journal} {Phys. Rev. Lett.}\ }\textbf {\bibinfo {volume} {120}},\
		\bibinfo {pages} {063604} (\bibinfo {year} {2018})}\BibitemShut {NoStop}%
	\bibitem [{\citenamefont {Brask}\ \emph {et~al.}(2015)\citenamefont {Brask},
		\citenamefont {Haack}, \citenamefont {Brunner},\ and\ \citenamefont
		{Huber}}]{Brask2015}%
	\BibitemOpen
	\bibfield  {author} {\bibinfo {author} {\bibfnamefont {J.~B.}\ \bibnamefont
			{Brask}}, \bibinfo {author} {\bibfnamefont {G.}~\bibnamefont {Haack}},
		\bibinfo {author} {\bibfnamefont {N.}~\bibnamefont {Brunner}},\ and\ \bibinfo
		{author} {\bibfnamefont {M.}~\bibnamefont {Huber}},\ }\bibfield  {title}
	{\bibinfo {title} {Autonomous quantum thermal machine for generating
			steady-state entanglement},\ }\href
	{https://doi.org/10.1088/1367-2630/17/11/113029} {\bibfield  {journal}
		{\bibinfo  {journal} {New J. Phys.}\ }\textbf {\bibinfo {volume} {17}},\
		\bibinfo {pages} {113029} (\bibinfo {year} {2015})}\BibitemShut {NoStop}%
	\bibitem [{\citenamefont {Khandelwal}\ \emph {et~al.}(2020)\citenamefont
		{Khandelwal}, \citenamefont {Palazzo}, \citenamefont {Brunner},\ and\
		\citenamefont {Haack}}]{Khandelwal2020}%
	\BibitemOpen
	\bibfield  {author} {\bibinfo {author} {\bibfnamefont {S.}~\bibnamefont
			{Khandelwal}}, \bibinfo {author} {\bibfnamefont {N.}~\bibnamefont {Palazzo}},
		\bibinfo {author} {\bibfnamefont {N.}~\bibnamefont {Brunner}},\ and\ \bibinfo
		{author} {\bibfnamefont {G.}~\bibnamefont {Haack}},\ }\bibfield  {title}
	{\bibinfo {title} {Critical heat current for operating an entanglement
			engine},\ }\href {https://doi.org/10.1088/1367-2630/ab9983} {\bibfield
		{journal} {\bibinfo  {journal} {New J. Phys.}\ }\textbf {\bibinfo {volume}
			{22}},\ \bibinfo {pages} {073039} (\bibinfo {year} {2020})}\BibitemShut
	{NoStop}%
	\bibitem [{\citenamefont {Bohr~Brask}\ \emph {et~al.}(2022)\citenamefont
		{Bohr~Brask}, \citenamefont {Clivaz}, \citenamefont {Haack},\ and\
		\citenamefont {Tavakoli}}]{Brask2022operational}%
	\BibitemOpen
	\bibfield  {author} {\bibinfo {author} {\bibfnamefont {J.}~\bibnamefont
			{Bohr~Brask}}, \bibinfo {author} {\bibfnamefont {F.}~\bibnamefont {Clivaz}},
		\bibinfo {author} {\bibfnamefont {G.}~\bibnamefont {Haack}},\ and\ \bibinfo
		{author} {\bibfnamefont {A.}~\bibnamefont {Tavakoli}},\ }\bibfield  {title}
	{\bibinfo {title} {Operational nonclassicality in minimal autonomous thermal
			machines},\ }\href {https://doi.org/10.22331/q-2022-03-22-672} {\bibfield
		{journal} {\bibinfo  {journal} {{Quantum}}\ }\textbf {\bibinfo {volume}
			{6}},\ \bibinfo {pages} {672} (\bibinfo {year} {2022})}\BibitemShut {NoStop}%
	\bibitem [{\citenamefont {Poulsen}\ and\ \citenamefont
		{Zinner}(2022)}]{Poulsen2022}%
	\BibitemOpen
	\bibfield  {author} {\bibinfo {author} {\bibfnamefont {K.}~\bibnamefont
			{Poulsen}}\ and\ \bibinfo {author} {\bibfnamefont {N.~T.}\ \bibnamefont
			{Zinner}},\ }\bibfield  {title} {\bibinfo {title} {Dark-state-induced heat
			rectification},\ }\href {https://doi.org/10.1103/PhysRevE.106.034116}
	{\bibfield  {journal} {\bibinfo  {journal} {Phys. Rev. E}\ }\textbf {\bibinfo
			{volume} {106}},\ \bibinfo {pages} {034116} (\bibinfo {year}
		{2022})}\BibitemShut {NoStop}%
	\bibitem [{\citenamefont {Poulsen}\ \emph {et~al.}(2022)\citenamefont
		{Poulsen}, \citenamefont {Santos},\ and\ \citenamefont
		{Zinner}}]{Poulsen2022b}%
	\BibitemOpen
	\bibfield  {author} {\bibinfo {author} {\bibfnamefont {K.}~\bibnamefont
			{Poulsen}}, \bibinfo {author} {\bibfnamefont {A.~C.}\ \bibnamefont
			{Santos}},\ and\ \bibinfo {author} {\bibfnamefont {N.~T.}\ \bibnamefont
			{Zinner}},\ }\bibfield  {title} {\bibinfo {title} {Quantum wheatstone
			bridge},\ }\href {https://doi.org/10.1103/PhysRevLett.128.240401} {\bibfield
		{journal} {\bibinfo  {journal} {Phys. Rev. Lett.}\ }\textbf {\bibinfo
			{volume} {128}},\ \bibinfo {pages} {240401} (\bibinfo {year}
		{2022})}\BibitemShut {NoStop}%
	\bibitem [{\citenamefont {Tavakoli}\ \emph {et~al.}(2018)\citenamefont
		{Tavakoli}, \citenamefont {Haack}, \citenamefont {Huber}, \citenamefont
		{Brunner},\ and\ \citenamefont {Brask}}]{Tavakoli2018}%
	\BibitemOpen
	\bibfield  {author} {\bibinfo {author} {\bibfnamefont {A.}~\bibnamefont
			{Tavakoli}}, \bibinfo {author} {\bibfnamefont {G.}~\bibnamefont {Haack}},
		\bibinfo {author} {\bibfnamefont {M.}~\bibnamefont {Huber}}, \bibinfo
		{author} {\bibfnamefont {N.}~\bibnamefont {Brunner}},\ and\ \bibinfo {author}
		{\bibfnamefont {J.~B.}\ \bibnamefont {Brask}},\ }\bibfield  {title} {\bibinfo
		{title} {Heralded generation of maximal entanglement in any dimension via
			incoherent coupling to thermal baths},\ }\href
	{https://doi.org/10.22331/q-2018-06-13-73} {\bibfield  {journal} {\bibinfo
			{journal} {{Quantum}}\ }\textbf {\bibinfo {volume} {2}},\ \bibinfo {pages}
		{73} (\bibinfo {year} {2018})}\BibitemShut {NoStop}%
	\bibitem [{\citenamefont {Tavakoli}\ \emph {et~al.}(2020)\citenamefont
		{Tavakoli}, \citenamefont {Haack}, \citenamefont {Brunner},\ and\
		\citenamefont {Brask}}]{Tavakoli2020}%
	\BibitemOpen
	\bibfield  {author} {\bibinfo {author} {\bibfnamefont {A.}~\bibnamefont
			{Tavakoli}}, \bibinfo {author} {\bibfnamefont {G.}~\bibnamefont {Haack}},
		\bibinfo {author} {\bibfnamefont {N.}~\bibnamefont {Brunner}},\ and\ \bibinfo
		{author} {\bibfnamefont {J.~B.}\ \bibnamefont {Brask}},\ }\bibfield  {title}
	{\bibinfo {title} {Autonomous multipartite entanglement engines},\ }\href
	{https://doi.org/10.1103/PhysRevA.101.012315} {\bibfield  {journal} {\bibinfo
			{journal} {Phys. Rev. A}\ }\textbf {\bibinfo {volume} {101}},\ \bibinfo
		{pages} {012315} (\bibinfo {year} {2020})}\BibitemShut {NoStop}%
	\bibitem [{\citenamefont {Diotallevi}\ \emph {et~al.}(2024)\citenamefont
		{Diotallevi}, \citenamefont {Annby-Andersson}, \citenamefont {Samuelsson},
		\citenamefont {Tavakoli},\ and\ \citenamefont
		{Bakhshinezhad}}]{Diotallevi2023}%
	\BibitemOpen
	\bibfield  {author} {\bibinfo {author} {\bibfnamefont {G.~F.}\ \bibnamefont
			{Diotallevi}}, \bibinfo {author} {\bibfnamefont {B.}~\bibnamefont
			{Annby-Andersson}}, \bibinfo {author} {\bibfnamefont {P.}~\bibnamefont
			{Samuelsson}}, \bibinfo {author} {\bibfnamefont {A.}~\bibnamefont
			{Tavakoli}},\ and\ \bibinfo {author} {\bibfnamefont {P.}~\bibnamefont
			{Bakhshinezhad}},\ }\bibfield  {title} {\bibinfo {title} {Steady-state
			entanglement production in a quantum thermal machine with continuous feedback
			control},\ }\href {https://doi.org/10.1088/1367-2630/ad3f3d} {\bibfield
		{journal} {\bibinfo  {journal} {New Journal of Physics}\ }\textbf {\bibinfo
			{volume} {26}},\ \bibinfo {pages} {053005} (\bibinfo {year}
		{2024})}\BibitemShut {NoStop}%
	\bibitem [{\citenamefont {de~Oliveira~Junior}\ \emph
		{et~al.}(2024)\citenamefont {de~Oliveira~Junior}, \citenamefont {Son},
		\citenamefont {Czartowski},\ and\ \citenamefont {Ng}}]{Oliveira2024}%
	\BibitemOpen
	\bibfield  {author} {\bibinfo {author} {\bibfnamefont {A.}~\bibnamefont
			{de~Oliveira~Junior}}, \bibinfo {author} {\bibfnamefont {J.}~\bibnamefont
			{Son}}, \bibinfo {author} {\bibfnamefont {J.}~\bibnamefont {Czartowski}},\
		and\ \bibinfo {author} {\bibfnamefont {N.~H.~Y.}\ \bibnamefont {Ng}},\
	}\bibfield  {title} {\bibinfo {title} {Entanglement generation from
			athermality},\ }\href {https://doi.org/10.1103/PhysRevResearch.6.033236}
	{\bibfield  {journal} {\bibinfo  {journal} {Phys. Rev. Res.}\ }\textbf
		{\bibinfo {volume} {6}},\ \bibinfo {pages} {033236} (\bibinfo {year}
		{2024})}\BibitemShut {NoStop}%
	\bibitem [{\citenamefont {Man}\ \emph {et~al.}(2019)\citenamefont {Man},
		\citenamefont {Tavakoli}, \citenamefont {Brask},\ and\ \citenamefont
		{Xia}}]{Man2019}%
	\BibitemOpen
	\bibfield  {author} {\bibinfo {author} {\bibfnamefont {Z.-X.}\ \bibnamefont
			{Man}}, \bibinfo {author} {\bibfnamefont {A.}~\bibnamefont {Tavakoli}},
		\bibinfo {author} {\bibfnamefont {J.~B.}\ \bibnamefont {Brask}},\ and\
		\bibinfo {author} {\bibfnamefont {Y.-J.}\ \bibnamefont {Xia}},\ }\bibfield
	{title} {\bibinfo {title} {Improving autonomous thermal entanglement
			generation using a common reservoir},\ }\href
	{https://doi.org/10.1088/1402-4896/ab0c51} {\bibfield  {journal} {\bibinfo
			{journal} {Phys. Scr.}\ }\textbf {\bibinfo {volume} {94}},\ \bibinfo {pages}
		{075101} (\bibinfo {year} {2019})}\BibitemShut {NoStop}%
	\bibitem [{\citenamefont {Prech}\ \emph {et~al.}(2023)\citenamefont {Prech},
		\citenamefont {Johansson}, \citenamefont {Nyholm}, \citenamefont {Landi},
		\citenamefont {Verdozzi}, \citenamefont {Samuelsson},\ and\ \citenamefont
		{Potts}}]{Prech2023}%
	\BibitemOpen
	\bibfield  {author} {\bibinfo {author} {\bibfnamefont {K.}~\bibnamefont
			{Prech}}, \bibinfo {author} {\bibfnamefont {P.}~\bibnamefont {Johansson}},
		\bibinfo {author} {\bibfnamefont {E.}~\bibnamefont {Nyholm}}, \bibinfo
		{author} {\bibfnamefont {G.~T.}\ \bibnamefont {Landi}}, \bibinfo {author}
		{\bibfnamefont {C.}~\bibnamefont {Verdozzi}}, \bibinfo {author}
		{\bibfnamefont {P.}~\bibnamefont {Samuelsson}},\ and\ \bibinfo {author}
		{\bibfnamefont {P.~P.}\ \bibnamefont {Potts}},\ }\bibfield  {title} {\bibinfo
		{title} {Entanglement and thermokinetic uncertainty relations in coherent
			mesoscopic transport},\ }\href
	{https://doi.org/10.1103/PhysRevResearch.5.023155} {\bibfield  {journal}
		{\bibinfo  {journal} {Phys. Rev. Res.}\ }\textbf {\bibinfo {volume} {5}},\
		\bibinfo {pages} {023155} (\bibinfo {year} {2023})}\BibitemShut {NoStop}%
	\bibitem [{\citenamefont {Lambropoulos}\ and\ \citenamefont
		{Petrosyan}(2007)}]{Lambropoulos2007}%
	\BibitemOpen
	\bibfield  {author} {\bibinfo {author} {\bibfnamefont {P.}~\bibnamefont
			{Lambropoulos}}\ and\ \bibinfo {author} {\bibfnamefont {D.}~\bibnamefont
			{Petrosyan}},\ }\href {https://doi.org/10.1007/978-3-540-34572-5/COVER}
	{\emph {\bibinfo {title} {{Fundamentals of quantum optics and quantum
					information}}}}\ (\bibinfo  {publisher} {Springer Berlin Heidelberg},\
	\bibinfo {year} {2007})\BibitemShut {NoStop}%
	\bibitem [{\citenamefont {Reimann}\ and\ \citenamefont
		{Manninen}(2002)}]{ReimannRMP2002}%
	\BibitemOpen
	\bibfield  {author} {\bibinfo {author} {\bibfnamefont {S.~M.}\ \bibnamefont
			{Reimann}}\ and\ \bibinfo {author} {\bibfnamefont {M.}~\bibnamefont
			{Manninen}},\ }\bibfield  {title} {\bibinfo {title} {Electronic structure of
			quantum dots},\ }\href {https://doi.org/10.1103/RevModPhys.74.1283}
	{\bibfield  {journal} {\bibinfo  {journal} {Rev. Mod. Phys.}\ }\textbf
		{\bibinfo {volume} {74}},\ \bibinfo {pages} {1283} (\bibinfo {year}
		{2002})}\BibitemShut {NoStop}%
	\bibitem [{\citenamefont {Gaudreau}\ \emph {et~al.}(2006)\citenamefont
		{Gaudreau}, \citenamefont {Studenikin}, \citenamefont {Sachrajda},
		\citenamefont {Zawadzki}, \citenamefont {Kam}, \citenamefont {Lapointe},
		\citenamefont {Korkusinski},\ and\ \citenamefont
		{Hawrylak}}]{GaudreauPRL2006}%
	\BibitemOpen
	\bibfield  {author} {\bibinfo {author} {\bibfnamefont {L.}~\bibnamefont
			{Gaudreau}}, \bibinfo {author} {\bibfnamefont {S.~A.}\ \bibnamefont
			{Studenikin}}, \bibinfo {author} {\bibfnamefont {A.~S.}\ \bibnamefont
			{Sachrajda}}, \bibinfo {author} {\bibfnamefont {P.}~\bibnamefont {Zawadzki}},
		\bibinfo {author} {\bibfnamefont {A.}~\bibnamefont {Kam}}, \bibinfo {author}
		{\bibfnamefont {J.}~\bibnamefont {Lapointe}}, \bibinfo {author}
		{\bibfnamefont {M.}~\bibnamefont {Korkusinski}},\ and\ \bibinfo {author}
		{\bibfnamefont {P.}~\bibnamefont {Hawrylak}},\ }\bibfield  {title} {\bibinfo
		{title} {Stability diagram of a few-electron triple dot},\ }\href
	{https://doi.org/10.1103/PhysRevLett.97.036807} {\bibfield  {journal}
		{\bibinfo  {journal} {Phys. Rev. Lett.}\ }\textbf {\bibinfo {volume} {97}},\
		\bibinfo {pages} {036807} (\bibinfo {year} {2006})}\BibitemShut {NoStop}%
	\bibitem [{\citenamefont {Rogge}\ and\ \citenamefont
		{Haug}(2009)}]{RoggeNJP2009}%
	\BibitemOpen
	\bibfield  {author} {\bibinfo {author} {\bibfnamefont {M.~C.}\ \bibnamefont
			{Rogge}}\ and\ \bibinfo {author} {\bibfnamefont {R.~J.}\ \bibnamefont
			{Haug}},\ }\bibfield  {title} {\bibinfo {title} {The three dimensionality of
			triple quantum dot stability diagrams},\ }\href
	{https://doi.org/10.1088/1367-2630/11/11/113037} {\bibfield  {journal}
		{\bibinfo  {journal} {New Journal of Physics}\ }\textbf {\bibinfo {volume}
			{11}},\ \bibinfo {pages} {113037} (\bibinfo {year} {2009})}\BibitemShut
	{NoStop}%
	\bibitem [{\citenamefont {Amaha}\ \emph {et~al.}(2008)\citenamefont {Amaha},
		\citenamefont {Hatano}, \citenamefont {Kubo}, \citenamefont {Tokura},
		\citenamefont {Guy~Austing},\ and\ \citenamefont
		{Tarucha}}]{AmahaPhysicaE2008}%
	\BibitemOpen
	\bibfield  {author} {\bibinfo {author} {\bibfnamefont {S.}~\bibnamefont
			{Amaha}}, \bibinfo {author} {\bibfnamefont {T.}~\bibnamefont {Hatano}},
		\bibinfo {author} {\bibfnamefont {T.}~\bibnamefont {Kubo}}, \bibinfo {author}
		{\bibfnamefont {Y.}~\bibnamefont {Tokura}}, \bibinfo {author} {\bibfnamefont
			{D.}~\bibnamefont {Guy~Austing}},\ and\ \bibinfo {author} {\bibfnamefont
			{S.}~\bibnamefont {Tarucha}},\ }\bibfield  {title} {\bibinfo {title}
		{Fabrication and characterization of a laterally coupled vertical triple
			quantum dot device},\ }\href {https://doi.org/10.1016/j.physe.2007.09.205}
	{\bibfield  {journal} {\bibinfo  {journal} {Physica E: Low-Dimensional
				Systems and Nanostructures}\ }\textbf {\bibinfo {volume} {40}},\ \bibinfo
		{pages} {1322 – 1324} (\bibinfo {year} {2008})}\BibitemShut {NoStop}%
	\bibitem [{\citenamefont {Bischoff}\ \emph {et~al.}(2013)\citenamefont
		{Bischoff}, \citenamefont {Varlet}, \citenamefont {Simonet}, \citenamefont
		{Ihn},\ and\ \citenamefont {Ensslin}}]{BischoffNJP2013}%
	\BibitemOpen
	\bibfield  {author} {\bibinfo {author} {\bibfnamefont {D.}~\bibnamefont
			{Bischoff}}, \bibinfo {author} {\bibfnamefont {A.}~\bibnamefont {Varlet}},
		\bibinfo {author} {\bibfnamefont {P.}~\bibnamefont {Simonet}}, \bibinfo
		{author} {\bibfnamefont {T.}~\bibnamefont {Ihn}},\ and\ \bibinfo {author}
		{\bibfnamefont {K.}~\bibnamefont {Ensslin}},\ }\bibfield  {title} {\bibinfo
		{title} {Electronic triple-dot transport through a bilayer graphene island
			with ultrasmall constrictions},\ }\href
	{https://doi.org/10.1088/1367-2630/15/8/083029} {\bibfield  {journal}
		{\bibinfo  {journal} {New Journal of Physics}\ }\textbf {\bibinfo {volume}
			{15}},\ \bibinfo {pages} {083029} (\bibinfo {year} {2013})}\BibitemShut
	{NoStop}%
	\bibitem [{\citenamefont {Brantut}\ \emph {et~al.}(2012)\citenamefont
		{Brantut}, \citenamefont {Meineke}, \citenamefont {Stadler}, \citenamefont
		{Krinner},\ and\ \citenamefont {Esslinger}}]{BrantutScience2012}%
	\BibitemOpen
	\bibfield  {author} {\bibinfo {author} {\bibfnamefont {J.-P.}\ \bibnamefont
			{Brantut}}, \bibinfo {author} {\bibfnamefont {J.}~\bibnamefont {Meineke}},
		\bibinfo {author} {\bibfnamefont {D.}~\bibnamefont {Stadler}}, \bibinfo
		{author} {\bibfnamefont {S.}~\bibnamefont {Krinner}},\ and\ \bibinfo {author}
		{\bibfnamefont {T.}~\bibnamefont {Esslinger}},\ }\bibfield  {title} {\bibinfo
		{title} {Conduction of ultracold fermions through a mesoscopic channel},\
	}\href {https://doi.org/10.1126/science.1223175} {\bibfield  {journal}
		{\bibinfo  {journal} {Science}\ }\textbf {\bibinfo {volume} {337}},\ \bibinfo
		{pages} {1069} (\bibinfo {year} {2012})}\BibitemShut {NoStop}%
	\bibitem [{\citenamefont {Potts}\ \emph {et~al.}(2021)\citenamefont {Potts},
		\citenamefont {Kalaee},\ and\ \citenamefont {Wacker}}]{Potts2021}%
	\BibitemOpen
	\bibfield  {author} {\bibinfo {author} {\bibfnamefont {P.~P.}\ \bibnamefont
			{Potts}}, \bibinfo {author} {\bibfnamefont {A.~A.~S.}\ \bibnamefont
			{Kalaee}},\ and\ \bibinfo {author} {\bibfnamefont {A.}~\bibnamefont
			{Wacker}},\ }\bibfield  {title} {\bibinfo {title} {A thermodynamically
			consistent markovian master equation beyond the secular approximation},\
	}\href {https://doi.org/10.1088/1367-2630/ac3b2f} {\bibfield  {journal}
		{\bibinfo  {journal} {New J. Phys.}\ }\textbf {\bibinfo {volume} {23}},\
		\bibinfo {pages} {123013} (\bibinfo {year} {2021})}\BibitemShut {NoStop}%
	\bibitem [{\citenamefont {Hofer}\ \emph {et~al.}(2017)\citenamefont {Hofer},
		\citenamefont {Perarnau-Llobet}, \citenamefont {Miranda}, \citenamefont
		{Haack}, \citenamefont {Silva}, \citenamefont {Brask},\ and\ \citenamefont
		{Brunner}}]{Hofer2017}%
	\BibitemOpen
	\bibfield  {author} {\bibinfo {author} {\bibfnamefont {P.~P.}\ \bibnamefont
			{Hofer}}, \bibinfo {author} {\bibfnamefont {M.}~\bibnamefont
			{Perarnau-Llobet}}, \bibinfo {author} {\bibfnamefont {L.~D.~M.}\ \bibnamefont
			{Miranda}}, \bibinfo {author} {\bibfnamefont {G.}~\bibnamefont {Haack}},
		\bibinfo {author} {\bibfnamefont {R.}~\bibnamefont {Silva}}, \bibinfo
		{author} {\bibfnamefont {J.~B.}\ \bibnamefont {Brask}},\ and\ \bibinfo
		{author} {\bibfnamefont {N.}~\bibnamefont {Brunner}},\ }\bibfield  {title}
	{\bibinfo {title} {{Markovian master equations for quantum thermal machines:
				local versus global approach}},\ }\href
	{https://doi.org/10.1088/1367-2630/aa964f} {\bibfield  {journal} {\bibinfo
			{journal} {New J. Phys.}\ }\textbf {\bibinfo {volume} {19}},\ \bibinfo
		{pages} {123037} (\bibinfo {year} {2017})}\BibitemShut {NoStop}%
	\bibitem [{foo({\natexlab{a}})}]{footnote}%
	\BibitemOpen
	\href@noop {} {\bibinfo {title} {Small variations in $\mu_3$ have no impact
			on the generated entanglement in the ideal parameter regime and have
			negligible impact on entanglement when perturbed away from the ideal
			parameter regime.}} %
	\bibitem [{foo({\natexlab{b}})}]{footnote2}%
	\BibitemOpen
	\href@noop {} {\bibinfo {title} {Note that one does not need to engineer the
			couplings to surpress the transitions corresponding to the other jump
			operators. in the considered limit; they become irrelevant regardless of
			their coupling strength.}} %
	\bibitem [{foo({\natexlab{c}})}]{footnote4}%
	\BibitemOpen
	\href@noop {} {\bibinfo {title} {In principle, a complex phase can be added
			in this superposition by using a magnetic field, through the peierls
			subsitution method \cite{Peierls1933}.}} %
	\bibitem [{\citenamefont {Yamamoto}(2005)}]{Yamamoto2005}%
	\BibitemOpen
	\bibfield  {author} {\bibinfo {author} {\bibfnamefont {N.}~\bibnamefont
			{Yamamoto}},\ }\bibfield  {title} {\bibinfo {title} {Parametrization of the
			feedback hamiltonian realizing a pure steady state},\ }\href
	{https://doi.org/10.1103/PhysRevA.72.024104} {\bibfield  {journal} {\bibinfo
			{journal} {Phys. Rev. A}\ }\textbf {\bibinfo {volume} {72}},\ \bibinfo
		{pages} {024104} (\bibinfo {year} {2005})}\BibitemShut {NoStop}%
	\bibitem [{\citenamefont {Kraus}\ \emph {et~al.}(2008)\citenamefont {Kraus},
		\citenamefont {B\"uchler}, \citenamefont {Diehl}, \citenamefont {Kantian},
		\citenamefont {Micheli},\ and\ \citenamefont {Zoller}}]{Kraus2008}%
	\BibitemOpen
	\bibfield  {author} {\bibinfo {author} {\bibfnamefont {B.}~\bibnamefont
			{Kraus}}, \bibinfo {author} {\bibfnamefont {H.~P.}\ \bibnamefont
			{B\"uchler}}, \bibinfo {author} {\bibfnamefont {S.}~\bibnamefont {Diehl}},
		\bibinfo {author} {\bibfnamefont {A.}~\bibnamefont {Kantian}}, \bibinfo
		{author} {\bibfnamefont {A.}~\bibnamefont {Micheli}},\ and\ \bibinfo {author}
		{\bibfnamefont {P.}~\bibnamefont {Zoller}},\ }\bibfield  {title} {\bibinfo
		{title} {Preparation of entangled states by quantum markov processes},\
	}\href {https://doi.org/10.1103/PhysRevA.78.042307} {\bibfield  {journal}
		{\bibinfo  {journal} {Phys. Rev. A}\ }\textbf {\bibinfo {volume} {78}},\
		\bibinfo {pages} {042307} (\bibinfo {year} {2008})}\BibitemShut {NoStop}%
	\bibitem [{\citenamefont {Albert}(2018)}]{VVAlbert}%
	\BibitemOpen
	\bibfield  {author} {\bibinfo {author} {\bibfnamefont {V.~V.}\ \bibnamefont
			{Albert}},\ }\bibfield  {title} {\bibinfo {title} {Lindbladians with multiple
			steady states: theory and applications},\ }\href
	{https://doi.org/10.48550/arXiv.1802.00010} {\bibfield  {journal} {\bibinfo
			{journal} {arXiv:1802.00010}\ } (\bibinfo {year} {2018})}\BibitemShut
	{NoStop}%
	\bibitem [{\citenamefont {Potts}(2019)}]{Pottsnotes}%
	\BibitemOpen
	\bibfield  {author} {\bibinfo {author} {\bibfnamefont {P.~P.}\ \bibnamefont
			{Potts}},\ }\href {https://doi.org/10.48550/arXiv.1906.07439} {\bibinfo
		{title} {Introduction to quantum thermodynamics (lecture notes)}} (\bibinfo
	{year} {2019})\BibitemShut {NoStop}%
	\bibitem [{\citenamefont {Emary}(2007)}]{EmaryPRB2007}%
	\BibitemOpen
	\bibfield  {author} {\bibinfo {author} {\bibfnamefont {C.}~\bibnamefont
			{Emary}},\ }\bibfield  {title} {\bibinfo {title} {Dark states in the
			magnetotransport through triple quantum dots},\ }\href
	{https://doi.org/10.1103/PhysRevB.76.245319} {\bibfield  {journal} {\bibinfo
			{journal} {Phys. Rev. B}\ }\textbf {\bibinfo {volume} {76}},\ \bibinfo
		{pages} {245319} (\bibinfo {year} {2007})}\BibitemShut {NoStop}%
	\bibitem [{\citenamefont {Wrze\ifmmode~\acute{s}\else \'{s}\fi{}niewski}\ and\
		\citenamefont {Weymann}(2018)}]{WrzesniewskiPRB2018}%
	\BibitemOpen
	\bibfield  {author} {\bibinfo {author} {\bibfnamefont {K.}~\bibnamefont
			{Wrze\ifmmode~\acute{s}\else \'{s}\fi{}niewski}}\ and\ \bibinfo {author}
		{\bibfnamefont {I.}~\bibnamefont {Weymann}},\ }\bibfield  {title} {\bibinfo
		{title} {Dark states in spin-polarized transport through triple quantum dot
			molecules},\ }\href {https://doi.org/10.1103/PhysRevB.97.075425} {\bibfield
		{journal} {\bibinfo  {journal} {Phys. Rev. B}\ }\textbf {\bibinfo {volume}
			{97}},\ \bibinfo {pages} {075425} (\bibinfo {year} {2018})}\BibitemShut
	{NoStop}%
	\bibitem [{foo({\natexlab{d}})}]{footnote3}%
	\BibitemOpen
	\href@noop {} {\bibinfo {title} {1 {GHz} is, for example, the relevant scale
			for state-of-the-art superconducting platforms \cite{Kjaergaard2019}. {For}
			semiconductor quantum dots, the scale can be one or two orders of magnitude
			larger.}} %
	\bibitem [{\citenamefont {Pedersen}\ and\ \citenamefont
		{Wacker}(2005)}]{Pedersen2005}%
	\BibitemOpen
	\bibfield  {author} {\bibinfo {author} {\bibfnamefont {J.~N.}\ \bibnamefont
			{Pedersen}}\ and\ \bibinfo {author} {\bibfnamefont {A.}~\bibnamefont
			{Wacker}},\ }\bibfield  {title} {\bibinfo {title} {Tunneling through
			nanosystems: Combining broadening with many-particle states},\ }\href
	{https://doi.org/10.1103/PhysRevB.72.195330} {\bibfield  {journal} {\bibinfo
			{journal} {Phys. Rev. B}\ }\textbf {\bibinfo {volume} {72}},\ \bibinfo
		{pages} {195330} (\bibinfo {year} {2005})}\BibitemShut {NoStop}%
	\bibitem [{\citenamefont {Pedersen}\ \emph {et~al.}(2007)\citenamefont
		{Pedersen}, \citenamefont {Lassen}, \citenamefont {Wacker},\ and\
		\citenamefont {Hettler}}]{Pedersen2007}%
	\BibitemOpen
	\bibfield  {author} {\bibinfo {author} {\bibfnamefont {J.~N.}\ \bibnamefont
			{Pedersen}}, \bibinfo {author} {\bibfnamefont {B.}~\bibnamefont {Lassen}},
		\bibinfo {author} {\bibfnamefont {A.}~\bibnamefont {Wacker}},\ and\ \bibinfo
		{author} {\bibfnamefont {M.~H.}\ \bibnamefont {Hettler}},\ }\bibfield
	{title} {\bibinfo {title} {Coherent transport through an interacting double
			quantum dot: Beyond sequential tunneling},\ }\href
	{https://doi.org/10.1103/PhysRevB.75.235314} {\bibfield  {journal} {\bibinfo
			{journal} {Phys. Rev. B}\ }\textbf {\bibinfo {volume} {75}},\ \bibinfo
		{pages} {235314} (\bibinfo {year} {2007})}\BibitemShut {NoStop}%
	\bibitem [{\citenamefont {Kir{\v{s}}anskas}\ \emph {et~al.}(2017)\citenamefont
		{Kir{\v{s}}anskas}, \citenamefont {Pedersen}, \citenamefont
		{Karlstr{\"{o}}m}, \citenamefont {Leijnse},\ and\ \citenamefont
		{Wacker}}]{KirsanskasComputPhysCommun2017}%
	\BibitemOpen
	\bibfield  {author} {\bibinfo {author} {\bibfnamefont {G.}~\bibnamefont
			{Kir{\v{s}}anskas}}, \bibinfo {author} {\bibfnamefont {J.~N.}\ \bibnamefont
			{Pedersen}}, \bibinfo {author} {\bibfnamefont {O.}~\bibnamefont
			{Karlstr{\"{o}}m}}, \bibinfo {author} {\bibfnamefont {M.}~\bibnamefont
			{Leijnse}},\ and\ \bibinfo {author} {\bibfnamefont {A.}~\bibnamefont
			{Wacker}},\ }\bibfield  {title} {\bibinfo {title} {Qme{Q} 1.0: An open-source
			{P}ython package for calculations of transport through quantum dot devices},\
	}\href {https://doi.org/10.1016/j.cpc.2017.07.024} {\bibfield  {journal}
		{\bibinfo  {journal} {Comput. Phys. Commun.}\ }\textbf {\bibinfo {volume}
			{221}},\ \bibinfo {pages} {317} (\bibinfo {year} {2017})}\BibitemShut
	{NoStop}%
	\bibitem [{\citenamefont {D\"ur}\ \emph {et~al.}(2000)\citenamefont {D\"ur},
		\citenamefont {Vidal},\ and\ \citenamefont {Cirac}}]{Dur2000}%
	\BibitemOpen
	\bibfield  {author} {\bibinfo {author} {\bibfnamefont {W.}~\bibnamefont
			{D\"ur}}, \bibinfo {author} {\bibfnamefont {G.}~\bibnamefont {Vidal}},\ and\
		\bibinfo {author} {\bibfnamefont {J.~I.}\ \bibnamefont {Cirac}},\ }\bibfield
	{title} {\bibinfo {title} {Three qubits can be entangled in two inequivalent
			ways},\ }\href {https://doi.org/10.1103/PhysRevA.62.062314} {\bibfield
		{journal} {\bibinfo  {journal} {Phys. Rev. A}\ }\textbf {\bibinfo {volume}
			{62}},\ \bibinfo {pages} {062314} (\bibinfo {year} {2000})}\BibitemShut
	{NoStop}%
	\bibitem [{\citenamefont {Cobucci}\ and\ \citenamefont
		{Tavakoli}(2024)}]{Cobucci2024}%
	\BibitemOpen
	\bibfield  {author} {\bibinfo {author} {\bibfnamefont {G.}~\bibnamefont
			{Cobucci}}\ and\ \bibinfo {author} {\bibfnamefont {A.}~\bibnamefont
			{Tavakoli}},\ }\bibfield  {title} {\bibinfo {title} {Detecting the
			dimensionality of genuine multiparticle entanglement},\ }\href
	{https://doi.org/10.1126/sciadv.adq4467} {\bibfield  {journal} {\bibinfo
			{journal} {Science Advances}\ }\textbf {\bibinfo {volume} {10}},\ \bibinfo
		{pages} {eadq4467} (\bibinfo {year} {2024})}\BibitemShut {NoStop}%
	\bibitem [{\citenamefont {Peierls}(1933)}]{Peierls1933}%
	\BibitemOpen
	\bibfield  {author} {\bibinfo {author} {\bibfnamefont {R.}~\bibnamefont
			{Peierls}},\ }\bibfield  {title} {\bibinfo {title} {Zur theorie des
			diamagnetismus von leitungselektronen},\ }\href
	{https://doi.org/10.1007/BF01342591} {\bibfield  {journal} {\bibinfo
			{journal} {Zeitschrift f{\"u}r Physik}\ }\textbf {\bibinfo {volume} {80}},\
		\bibinfo {pages} {763} (\bibinfo {year} {1933})}\BibitemShut {NoStop}%
	\bibitem [{\citenamefont {Kjaergaard}\ \emph {et~al.}(2020)\citenamefont
		{Kjaergaard}, \citenamefont {Schwartz}, \citenamefont {Braum\"{u}ller},
		\citenamefont {Krantz}, \citenamefont {Wang}, \citenamefont {Gustavsson},\
		and\ \citenamefont {Oliver}}]{Kjaergaard2019}%
	\BibitemOpen
	\bibfield  {author} {\bibinfo {author} {\bibfnamefont {M.}~\bibnamefont
			{Kjaergaard}}, \bibinfo {author} {\bibfnamefont {M.~E.}\ \bibnamefont
			{Schwartz}}, \bibinfo {author} {\bibfnamefont {J.}~\bibnamefont
			{Braum\"{u}ller}}, \bibinfo {author} {\bibfnamefont {P.}~\bibnamefont
			{Krantz}}, \bibinfo {author} {\bibfnamefont {J.~I.-J.}\ \bibnamefont {Wang}},
		\bibinfo {author} {\bibfnamefont {S.}~\bibnamefont {Gustavsson}},\ and\
		\bibinfo {author} {\bibfnamefont {W.~D.}\ \bibnamefont {Oliver}},\ }\bibfield
	{title} {\bibinfo {title} {Superconducting qubits: Current state of play},\
	}\href {https://doi.org/10.1146/annurev-conmatphys-031119-050605} {\bibfield
		{journal} {\bibinfo  {journal} {Annu. Rev. Condens. Matter Phys.}\ }\textbf
		{\bibinfo {volume} {11}},\ \bibinfo {pages} {369} (\bibinfo {year}
		{2020})}\BibitemShut {NoStop}%
	\bibitem [{\citenamefont {Eisenberg}\ \emph {et~al.}(2002)\citenamefont
		{Eisenberg}, \citenamefont {Held},\ and\ \citenamefont
		{Altshuler}}]{Eisenberg2002}%
	\BibitemOpen
	\bibfield  {author} {\bibinfo {author} {\bibfnamefont {E.}~\bibnamefont
			{Eisenberg}}, \bibinfo {author} {\bibfnamefont {K.}~\bibnamefont {Held}},\
		and\ \bibinfo {author} {\bibfnamefont {B.~L.}\ \bibnamefont {Altshuler}},\
	}\bibfield  {title} {\bibinfo {title} {Dephasing times in closed quantum
			dots},\ }\href {https://doi.org/10.1103/PhysRevLett.88.136801} {\bibfield
		{journal} {\bibinfo  {journal} {Phys. Rev. Lett.}\ }\textbf {\bibinfo
			{volume} {88}},\ \bibinfo {pages} {136801} (\bibinfo {year}
		{2002})}\BibitemShut {NoStop}%
	\bibitem [{\citenamefont {Cesari}\ \emph {et~al.}(2010)\citenamefont {Cesari},
		\citenamefont {Langbein},\ and\ \citenamefont {Borri}}]{Cesari2010}%
	\BibitemOpen
	\bibfield  {author} {\bibinfo {author} {\bibfnamefont {V.}~\bibnamefont
			{Cesari}}, \bibinfo {author} {\bibfnamefont {W.}~\bibnamefont {Langbein}},\
		and\ \bibinfo {author} {\bibfnamefont {P.}~\bibnamefont {Borri}},\ }\bibfield
	{title} {\bibinfo {title} {Dephasing of excitons and multiexcitons in
			undoped and $p$-doped inas/gaas quantum dots-in-a-well},\ }\href
	{https://doi.org/10.1103/PhysRevB.82.195314} {\bibfield  {journal} {\bibinfo
			{journal} {Phys. Rev. B}\ }\textbf {\bibinfo {volume} {82}},\ \bibinfo
		{pages} {195314} (\bibinfo {year} {2010})}\BibitemShut {NoStop}%
	\bibitem [{\citenamefont {Lindwall}\ \emph {et~al.}(2007)\citenamefont
		{Lindwall}, \citenamefont {Wacker}, \citenamefont {Weber},\ and\
		\citenamefont {Knorr}}]{Lindwall2007}%
	\BibitemOpen
	\bibfield  {author} {\bibinfo {author} {\bibfnamefont {G.}~\bibnamefont
			{Lindwall}}, \bibinfo {author} {\bibfnamefont {A.}~\bibnamefont {Wacker}},
		\bibinfo {author} {\bibfnamefont {C.}~\bibnamefont {Weber}},\ and\ \bibinfo
		{author} {\bibfnamefont {A.}~\bibnamefont {Knorr}},\ }\bibfield  {title}
	{\bibinfo {title} {Zero-phonon linewidth and phonon satellites in the optical
			absorption of nanowire-based quantum dots},\ }\href
	{https://doi.org/10.1103/PhysRevLett.99.087401} {\bibfield  {journal}
		{\bibinfo  {journal} {Phys. Rev. Lett.}\ }\textbf {\bibinfo {volume} {99}},\
		\bibinfo {pages} {087401} (\bibinfo {year} {2007})}\BibitemShut {NoStop}%
	\bibitem [{\citenamefont {Thorgrimsson}\ \emph {et~al.}(2017)\citenamefont
		{Thorgrimsson}, \citenamefont {Kim}, \citenamefont {Yang}, \citenamefont
		{Smith}, \citenamefont {Simmons}, \citenamefont {Ward}, \citenamefont
		{Foote}, \citenamefont {Corrigan}, \citenamefont {Savage}, \citenamefont
		{Lagally}, \citenamefont {Friesen}, \citenamefont {Coppersmith},\ and\
		\citenamefont {Eriksson}}]{Thorg2017}%
	\BibitemOpen
	\bibfield  {author} {\bibinfo {author} {\bibfnamefont {B.}~\bibnamefont
			{Thorgrimsson}}, \bibinfo {author} {\bibfnamefont {D.}~\bibnamefont {Kim}},
		\bibinfo {author} {\bibfnamefont {Y.-C.}\ \bibnamefont {Yang}}, \bibinfo
		{author} {\bibfnamefont {L.~W.}\ \bibnamefont {Smith}}, \bibinfo {author}
		{\bibfnamefont {C.~B.}\ \bibnamefont {Simmons}}, \bibinfo {author}
		{\bibfnamefont {D.~R.}\ \bibnamefont {Ward}}, \bibinfo {author}
		{\bibfnamefont {R.~H.}\ \bibnamefont {Foote}}, \bibinfo {author}
		{\bibfnamefont {J.}~\bibnamefont {Corrigan}}, \bibinfo {author}
		{\bibfnamefont {D.~E.}\ \bibnamefont {Savage}}, \bibinfo {author}
		{\bibfnamefont {M.~G.}\ \bibnamefont {Lagally}}, \bibinfo {author}
		{\bibfnamefont {M.}~\bibnamefont {Friesen}}, \bibinfo {author} {\bibfnamefont
			{S.~N.}\ \bibnamefont {Coppersmith}},\ and\ \bibinfo {author} {\bibfnamefont
			{M.~A.}\ \bibnamefont {Eriksson}},\ }\bibfield  {title} {\bibinfo {title}
		{Extending the coherence of a quantum dot hybrid qubit},\ }\href
	{https://doi.org/10.1038/s41534-017-0034-2} {\bibfield  {journal} {\bibinfo
			{journal} {npj Quantum Information}\ }\textbf {\bibinfo {volume} {3}},\
		\bibinfo {pages} {32} (\bibinfo {year} {2017})}\BibitemShut {NoStop}%
	\bibitem [{git()}]{github}%
	\BibitemOpen
	\href {https://github.com/gedaskir/qmeq} {\bibinfo  {journal}
		{https://github.com/gedaskir/qmeq}\ }\BibitemShut {NoStop}%
\end{thebibliography}

\appendix
\onecolumngrid

\section*{Appendix}

\section*{A. Machine to generate $\ket{\Psi^-}$:  Liouvillian and steady state}
\label{app:1}
\noindent The Liouvillian of the three-qubit machine, $\mathcal L$, can be represented in matrix form by vectorizing the three qubit state. Since the $8\times8$ three-qubit state transforms into a $64\times 1$ vector, $\mathcal L$ can be represented as a $64\times 64$ matrix. However, under the limits (3) of the main text, the doubly occupied subspace plays no role at long times and can simply be eliminated from the dynamics. Moreover, restricting to the steady-state subspace, the Liouvillian can be reduced to a $10\times10$ matrix. In the basis $\{ \ketbra{100}{100},\ketbra{100}{010},\ketbra{100}{001},\ketbra{010}{100},\ketbra{010}{010},\ketbra{010}{001},\ketbra{001}{100},\ketbra{001}{010},\ketbra{001}{001},\ketbra{000}{000}\}$, the Liouvillian expressed in matrix form is 
\begin{align}
	\ket{\mathcal L}\rangle = \left(
	\begin{array}{cccccccccc}
		0 & 0 & i g_{13} & 0 & 0 & 0 & -i g_{13} & 0 & 0 & \gamma_{100}^+\\
		0 & 0 & i g_{23} & 0 & 0 & 0 & 0 & -i g_{13} & 0 & 0 \\
		i g_{13} & i g_{23} & -\frac{\gamma_{300}^-}{2} & 0 & 0 & 0 & 0 & 0 & -i g_{13} & 0 \\
		0 & 0 & 0 & 0 & 0 & i g_{13} & -i g_{23} & 0 & 0 & 0 \\
		0 & 0 & 0 & 0 & 0 & i g_{23} & 0 & -i g_{23} & 0 & 0\\
		0 & 0 & 0 & i g_{13} & i g_{23} & -\frac{\gamma_{300}^-}{2} & 0 & 0 & -i g_{23} & 0 \\
		-i g_{13} & 0 & 0 & -i g_{23} & 0 & 0 & -\frac{\gamma_{300}^-}{2} & 0 & i g_{13} & 0 \\
		0 & -i g_{13} & 0 & 0 & -i g_{23} & 0 & 0 & -\frac{\gamma_{300}^-}{2} & i g_{23} & 0 \\
		0 & 0 & -i g_{13} & 0 & 0 & -i g_{23} & i g_{13} & i g_{23} & -\gamma_{300}^-& \gamma_{300}^+ \\
		0 & 0 & 0 & 0 & 0 & 0 & 0 & 0 & \gamma_{300}^- & -\gamma_{100}^+ -\gamma_{300}^+ \\
	\end{array}
	\right).
\end{align}The steady state $\ketbra{\Psi_{\text{ss}}}{\Psi_{\text{ss}}}$, with
\begin{align}\label{eq:ss1}
	\ket{\Psi_{\text{ss}} }= \left(\hspace*{-0.1cm}\frac{g_{23}}{\sqrt{g^2_{23}+g^2_{13}}}\ket{10}-\frac{g_{13}}{\sqrt{g^2_{23}+g^2_{13}}}\ket{01} \hspace*{-0.1cm}\right)\otimes \ket{0},
\end{align}
is the eigenstate corresponding to the zero eigenvalue of $\mathcal L$.

\section*{B. Scheme to generate $\ket{W_n}$: analytical solution}
\label{app:proof}
\begin{figure}
	\centering
	\includegraphics[width=1\textwidth]{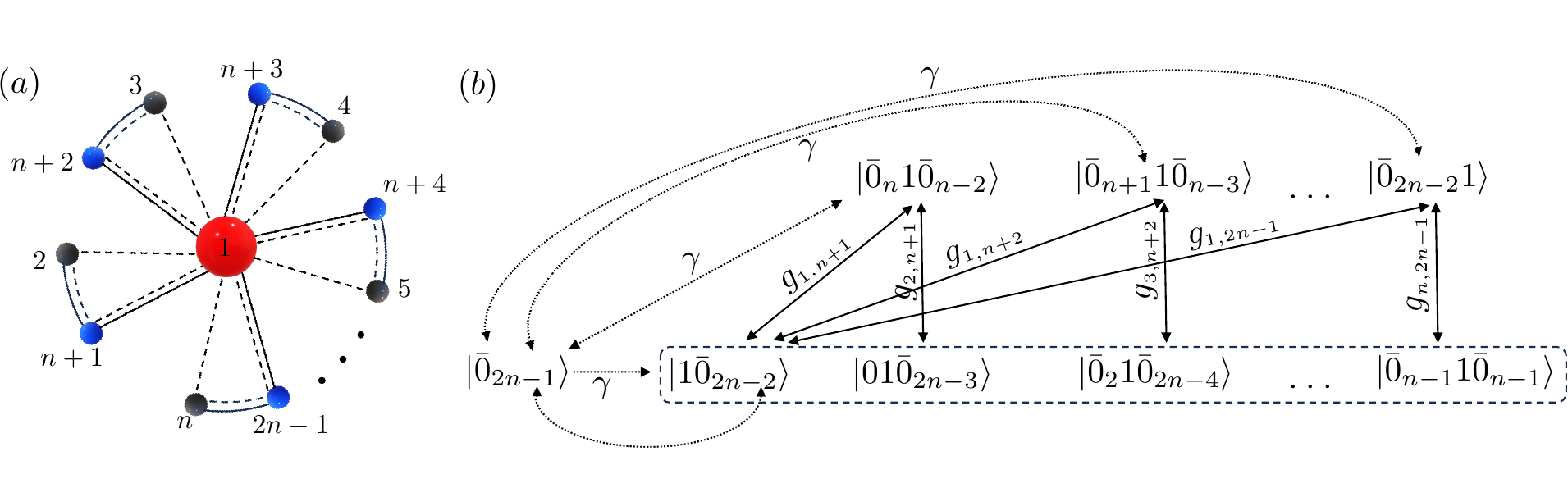}
	\caption{(a)  A two-dimensional depiction of the $(2n-1)$-qubit scheme for generating an $n$-qubit W state. The solid lines indicate the inter-qubit couplings, while the dashed ones indicate Coulomb interactions. The first qubit is coupled to all qubits $n+j$ with strength $g_{1,n+j}$ and the qubits $1+j$ are coupled to qubits  $n+j$ with strength $g_{1+j,n+j}$ ($j=1,2,...,n-1$) . The red qubit is coupled with the high-bias reservoir, while the blue qubits are coupled with the low-bias reservoirs (i.e., they are the sink qubits). In the steady state, a W-like state is created between the first $n$ qubits, while the remaining $n-1$ qubits are pushed into their ground states. (b) Generalized coherent population trapping. The state of the qubits numbered from 1 to $n$ is coherently trapped between $n$ single-particle states (shown in the box), while qubits numbered from $(n+1)$ to $(2n-1)$ are pushed into their ground state. The inter-qubit interactions are depicted as solid arrows and qubit-reservoir interactions as dotted arrows. We emphasize that the ideal regime, system-reservoir couplings play no role in the steady state and can be completely arbitrary and distinct from each other (while remaining within the validity of the master equation) for the realization of the scheme.}
	\label{fig:CPT_gen}
\end{figure}
\noindent Motivated by the three-qubit machine, we consider the generalised scheme discussed in the main text. The scheme was illustrated in Fig.~4 in the main text and a more elaborate  illustration is given in  Fig.~\ref{fig:CPT_gen}(a). The system consists of $2n-1$ qubits of degenerate energy $\varepsilon$. The first qubit is connected to a fermionic reservoir at temperature $T$, chemical potential $\mu$. The qubits $n+1$ to $2n-1$ are connected to reservoirs at temperature $T$. We refer to these qubits as sink qubits as they are connected to the low-bias reservoir (at zero chemical potential). The first qubit is interacting with qubits $n+j$ ($1\leq  j\leq n-1$)  with strength $g_{1,n+j}$ and the qubits $1+j$ are interacting with qubits $n+j$ with strength $g_{1+j,n+j}$, through flip-flop interactions. We further assume that all qubits have pairwise Coulomb-like interaction with strength $U$. This condition can be relaxed. In principle, the scheme requires Coulomb interactions only between the pairs $(1,n+j)$, $(1,1+j)$ and $(1+j,n+j)$. However, assuming all pairs to have such interactions will further simplify the proof of the scheme, as we will see later. For convenience, we divide the Hamiltonian of the setup into four parts. $H_1$ is the bare Hamiltonian of the qubits, given by,
\begin{align}
	H_1 =\varepsilon \sum_{i_m\in\{0,1\}} \left\{\left( \sum_{j=1}^{2n-1}i_j\right)\ketbra{i_1\,i_2\cdot\cdot\cdot i_{2n-1}}{i_1\,i_2\cdot\cdot\cdot i_{2n-1}}\right\}.
\end{align}$H_2$ is the Hamiltonian due to Coulomb interactions,
\begin{align}
	H_2 = \sum_{i_m\in\{0,1\}}\left\{\frac{1}{2}U\left(\sum_{j=1}^{2n-1}i_j\Big(\sum_{j=1}^{2n-1}i_j-1 \Big)\ketbra{i_1i_2\cdot\cdot\cdot i_{2n-1}}{i_1i_2\cdot\cdot\cdot i_{2n-1}}\right)\right\}.
\end{align}$H_3$ is the interaction between qubit 1 and the sink qubits $n+j$,
\begin{equation}
	\begin{aligned}
		H_3 = \sum_{j=1}^{n}\sum_{i_m=\{0,1\}}\left\{g_{1,n+j}\bigg( \ketbra{1\,i_2\,i_3\cdot\cdot\cdot i_{n+j-1}0\cdot\cdot\cdot i_{2n-1}}{0\,i_2\,i_3\cdot\cdot\cdot i_{n+j-1}1\cdot\cdot\cdot i_{2n-1}}+\text{h.c.} \bigg)\right\}.
	\end{aligned}
\end{equation}Lastly, $H_4$ is the interaction between qubits $1+j$ and the sink qubits $n+j$,
\begin{equation}
	\hspace*{-0.5cm}H_4 = \sum_{j=1}^{n}\sum_{i_m=\{0,1\}}\hspace*{-0.4cm}g_{1+j,n+j} \left(\ketbra{i_1\,i_2\cdot\cdot\cdot i_{j}\,1\,i_{j+2}\cdot\cdot\cdot i_{n+j-1}\,0\,i_{n+j+1} \cdot\cdot\cdot i_{2n-1}}{i_1\,i_2\cdot\cdot\cdot i_{j}\,0\,i_{j+2}\cdot\cdot\cdot i_{n+j-1}\,1\,i_{n+j+1} \cdot\cdot\cdot i_{2n-1}} +\text{h.c.}\right)
\end{equation}The total Hamiltonian is then $H=\sum_{j=1}^{4}H_j$. We now take the limits $U\to\infty$, $\mu\to\infty$ and $U/\mu\to\infty$. For practical purposes, these can be relaxed to $U\gg T$, $\mu\gg T$ and $U\gg\mu$. These conditions have the following consequences on the evolution,
\begin{enumerate}
	\item{When there is no excitation in the system (i.e., the system is in the ground state), the high-bias reservoir can only transfer excitations into the system.}
	\item{When there is already an excitation in the system, the high-bias reservoir cannot overcome the high Coulomb repulsion to transfer another excitation.}
	\item{The excitations can only leave through the low-bias reservoirs.}
\end{enumerate}The above properties ensure that the doubly occupied subspace of the $2n-1$-qubit state is completely eliminated and the transport has a unidirectional property (excitations from the high-bias reservoir travel into the low-bias reservoir). We then have the following jump operators,
\begin{equation}
	\begin{aligned}
		&L_{1,+} = \ketbra{1\bar0_{2n-2}}{\bar0_{2n-1}} \\
		&L_{1,-} = \ketbra{0\,i_2\cdot\cdot\cdot i_{2n-1}}{1\,i_2\cdot\cdot\cdot i_{2n-1}}\\
		&L_{p,-} =\ketbra{i_1\,i_2\cdot\cdot\cdot i_{p-1}\,0\,i_{p+1}\cdot\cdot\cdot i_{2n-1}}{i_1\,i_2\cdot\cdot\cdot i_{p-1}\,1\,i_{p+1}\cdot\cdot\cdot i_{2n-1}}  \\
		&L_{p,+} =\ketbra{i_1\,i_2\cdot\cdot\cdot i_{p-1}\,1\,i_{p+1}\cdot\cdot\cdot i_{2n-1}}{i_1\,i_2\cdot\cdot\cdot i_{p-1}\,0\,i_{p+1}\cdot\cdot\cdot i_{2n-1}}     ,
	\end{aligned}
\end{equation}where $p\in\{n+1,n+2,\cdot\cdot\cdot,2n-1\}$,  $i_m\in\{0,1\}$ and $\bar 0_{k}\coloneqq \ket{0}\otimes\ket{0}\otimes\cdot\cdot\cdot\otimes\ket{0}$ is the ground state of $k$ qubits. These are the only possible transitions within the setup. We further assume that the initial state of the system is at most singly excited. This allows us to keep only the operators for which all $i_m$s are zero. The general case can be treated in a similar, albeit more tedious manner, and leads to the same steady state. We therefore have the Lindblad equation,
\begin{align}
	\dot\rho = -i\left[H,\rho \right] + \gamma_1^+\mathcal D\left[L_{1,+}\right]\rho +\gamma_1^- \mathcal D\left[L_{1,-}\right]\rho + \gamma_p^- \mathcal D\left[L_{p,-}\right]\rho  + \gamma_p^+ \mathcal D\left[L_{p,+}\right]\rho,
\end{align}where $\mathcal D[A]\rho\coloneqq A\rho A^\dagger - \left\{A^\dagger A,\rho\right\}/2$. The rates are determined by the Fermi-Dirac distribution of the reservoirs, $\gamma_j^+ = \gamma_j n_F\left(\varepsilon,\mu,T \right)$ and $\gamma_j^- = \gamma_j \left( 1-n_F\left(\varepsilon,\mu,T \right)\right)$, with $n_F\left(\varepsilon,\mu,T \right) = 1/\left( e^{(\varepsilon-\mu)/T}+1\right)$. With the above mentioned assumptions, we have that
\begin{equation}
	\begin{aligned}
		\gamma_1^+ =\gamma_1, \,\,\text{and}\,\, \gamma_1^- = 0.
	\end{aligned}
\end{equation}We finally have the Lindblad equation,
\begin{align}
	\dot\rho \coloneqq \mathcal L_n\rho= -i\left[H,\rho \right] + \gamma_1\mathcal D\left[L_{1,+}\right]\rho  + \gamma_p^- \mathcal D\left[L_{p,-}\right]\rho + \gamma_p^+ \mathcal D\left[L_{p,+}\right]\rho.
\end{align}The steady state of the system satisfies $\mathcal L_n\rho_{\text{ss}}^n=0$.
For arbitrary $n$ and under the limits (3) in the main text, the machine in Fig. 4 of the main text produces a pure, partially entangled W-like steady state between qubits $1$-$n$, while the remaining qubits are pushed to their ground state.
\begin{align}\label{eq:genss1}
	\ket{\Psi^n_\text{ss}} = \frac{1}{\sqrt{\beta}}\left( \ket{10\cdot\cdot\cdot0}-\alpha_1\ket{01\cdot\cdot\cdot0}-\alpha_2\ket{001\cdot\cdot\cdot0}-\cdot\cdot\cdot-\alpha_{n-1}\ket{0\cdot\cdot\cdot01}\right)\otimes\ket{\bar 0_{n-1}},
\end{align}where,
\begin{align}
	\alpha_j = \frac{g_{1,n+j}}{g_{1+j,n+j}}\quad\text{and}\quad \beta =1+\sum_{j=1}^{n-1}\alpha_j^2.
\end{align}For $n=2$, the expression obtained can be rewritten as Eq. \eqref{eq:ss1}. The above result can be understood as a generalization of coherent population trapping, as depicted in Fig. \ref{fig:CPT_gen} (b). In the coherent population trapping explained in the main text, a two-qubit state is coherently trapped between two single-particle states. Here, the state of qubits 1-$n$ is trapped between $n$ single-particle states. Clearly, if all $\alpha_j=1$, we obtain a W state between the first $n$ qubits and the steady state takes the form,
\begin{align}
	\ket{W_n}=\frac{1}{\sqrt{n}}\hspace*{-0.1cm}\left(\ket{100\ldots 0}-\ket{010\ldots 0 }-\ket{000\ldots 1}\right)\otimes \ket{\bar{0}_{n-1}}.
\end{align}We denote the corresponding density matrix by $\rho_{\tiny{W_n}}\coloneqq \ketbra{W_n}{W_n}$.

\subsection{Proof of steady state}
\noindent 
We now show that $\rho_{\tiny{W_n}}$ is the steady state of the general scheme, i.e., that $\mathcal L_n\rho_{W_{n}}=0$, when certain conditions are satisfied between the couplings.
\subsubsection{Dissipators}
\begin{enumerate}
	\item The only dissipator corresponding to qubit 1 is  $L_{1,+}=\ketbra{1\bar 0_{2n-2}}{\bar 0_{2n-1}}$. We have that $L_{1,+}\ket{W_n}=0$ because it involves the overlap between a ground state $\ket{\bar0_{2n-1}}$ with $\ket{W_n}$ which is a superposition of singly occupied states.
	
	\item The first dissipator for qubit $p$, $L_{p,+}=\ketbra{\bar0_{p-1}1\bar0_{2n-p-1}}{\bar0_{2n-1}}$, satisfies $L_{p,+}\ket{W_n} =0$ for the same reason as above.
	
	\item The second dissipator for qubit $p$ is $L_{p,-} = \ketbra{\bar0_{2n-1}}{\bar0_{p-1}1\bar0_{2n-p-1}}$. Note that in $\ket{\bar0_{p-1}1\bar0_{2n-p-1}}$, the occupation lies in the sink subspace (as $p>n$). On the other hand, the sink subspace of $\ket{W_n}$ is completely in the ground state. Therefore, we have that $L_{p,-}\ket{W_n}=0$.
\end{enumerate}
The above observations imply that  the state $\ket{W_n}$ is unaffected by dissipation. Mathematically, 
\begin{align}
	\mathcal D[L_{1,+}]\rho_{W_n} =  	\mathcal D[L_{p,+}]\rho_{W_n} =	\mathcal D[L_{p,-}]\rho_{W_n} =0.
\end{align}

\subsubsection{Hamiltonian}
\noindent It can be easily shown that in the commutator $[H,\rho_{W_n}]$, $H_1$ and $H_2$ have zero contribution. Therefore, we focus on $H_3$ and $H_4$. Here, only the terms with at most single occupation in the Hamiltonian have a non-zero contribution in the commutator $[H,\rho]$. Therefore, without loss of generality, we can eliminate all other terms and write the Hamiltonian as,
\begin{equation}
	\begin{aligned}
		&H_3^{(1)} = \sum_{j=1}^{n-1} g_{1,n+j} \ketbra{1\bar0_{2n-2}}{\bar 0_{n+j-1}1\bar0_{n-j-1}}, \quad  H_3^{(2)} = H_3^{(1)\dagger}\\
		&H_4^{(1)} = \sum_{j=1}^{n-1} g_{1+j,n+j} \ketbra{\bar 0_{j}1\bar 0_{2n-j-2}}{\bar 0_{n+j-1}1\bar 0_{n-j-1}}
		, \quad  H_3^{(2)} = H_3^{(1)\dagger},
	\end{aligned}
\end{equation}with $H_j=H_j^{(1)}+H_j^{(2)}$ ($j=3,4$).The terms in the commutator  can now be treated separately. First,
For the term, $H_3^{(1)}\rho_{W_n}$, we note that the sink subspace in $\ket{W_n}$ is unoccupied (i.e., the sink qubits are in the ground state). Evidently, its overlap with $\ket{\bar 0_{n+j-1}1\bar0_{n-j-1}}$ (which has a singly occupied sink subspace) is necessarily zero, implying that $H_{3}^{(1)}\rho_{W_n}=0$. For the same reason, $\rho_{W_n}H_3^{(2)}=0$. Overall, the contribution of $H_3$ takes the form,
\begin{align}
	[H_{3},\rho_{W_n}] &= H_3^{(2)}\rho_{W_n} - \rho_{W_n}H_3^{(1)}\nonumber \\ 
	&=\sum_{i=1}^{n}\sum_{j=1}^{n-1}g_{1,n+j}\left(\ketbra{\bar0_{i-1}1\bar0_{2n-i-1}}{\bar0_{n+j-1}1\bar0_{n-j-1}} -  \ketbra{\bar0_{n+j-1}1\bar0_{n-j-1}}{\bar0_{i-1}1\bar0_{2n-i-1}}\right).
\end{align}	In a similar manner, it can be shown that,
\begin{align}
	[H_{4},\rho_{W_n}] &=  H_4^{(2)}\rho_{W_n}-\rho_{W_n}H_4^{(1)}  \nonumber \\ 
	&=\sum_{i=1}^{n}\sum_{j=1}^{n-1}g_{1+j,n+j}\left(\ketbra{\bar0_{n+j-1}1\bar0_{n-j-1}}{\bar0_{i-1}1\bar0_{2n-i-1}} - \ketbra{\bar0_{i-1}1\bar0_{2n-i-1}}{\bar0_{n+j-1}1\bar0_{n-j-1}} \right). 
\end{align}	Finally, the full commutator simplifies to
\begin{align}
	[H,\rho_{W_n}] = \sum_{i=1}^{n}\sum_{j=1}^{n-1}\left( g_{1+j,n+j}-g_{1,n+j}\right)\left(\ketbra{\bar0_{n+j-1}1\bar0_{n-j-1}}{\bar0_{i-1}1\bar0_{2n-i-1}} - \ketbra{\bar0_{i-1}1\bar0_{2n-i-1}}{\bar0_{n+j-1}1\bar0_{n-j-1}} \right). 
\end{align}Overall, we have that $[H,\rho_{W_n}] =0$, when the couplings satisfy $g_{1,n+j} = g_{1+j,n+j}$. Therefore, when this condition is true, we have $\mathcal L\ketbra{W_n}{W_n}=0$. In other words, $\ket{W_n}$ is the steady state of the Liouvillian $\mathcal L_n$. More generally, it can be proven on similar lines that when this condition is not satisfied, the pure and partially entangled state $\ket{\Psi_{\text{ss}}^n}$, given in Eq. \eqref{eq:genss1}, is the steady state.

\section*{C. Robustness to dephasing and system-reservoir couplings}
In the main text, we have not commented upon the possible dephasing of the qubits coupled to the fermionic reservoirs. By depleting the coherence, dephasing is evidently detrimental for entanglement generation with our machine. In Fig. \ref{fig:dephasing} (a), we plot the steady state fidelity as a function of $\gamma_z/\gamma$, where $\gamma_z$ is the dephasing rate of qubits 1 and 3.  Under $\gamma_z/\gamma\sim1$ \%, which is within experimental reach \cite{Eisenberg2002, Cesari2010, Lindwall2007,Thorg2017}, the scheme can still achieve above 90\% fidelity.\par
In the main text, we have discussed that in the ideal limit of our scheme, the system-reservoir couplings play no role in the steady state. However, this is not true in general outside of this regime, for example, where $U$ and $\mu$ are not large enough. In Fig. \ref{fig:dephasing} (b), we further show the impact of a mismatch between sink and source couplings with a  constant large $U$ and varying $\mu$. We find that for $U$ and $\mu$ approaching a suitable regime for our scheme, the mismatch leads to negligible change in the steady-state fidelity. 
\begin{figure}
	\centering
	\includegraphics[width=0.8\textwidth]{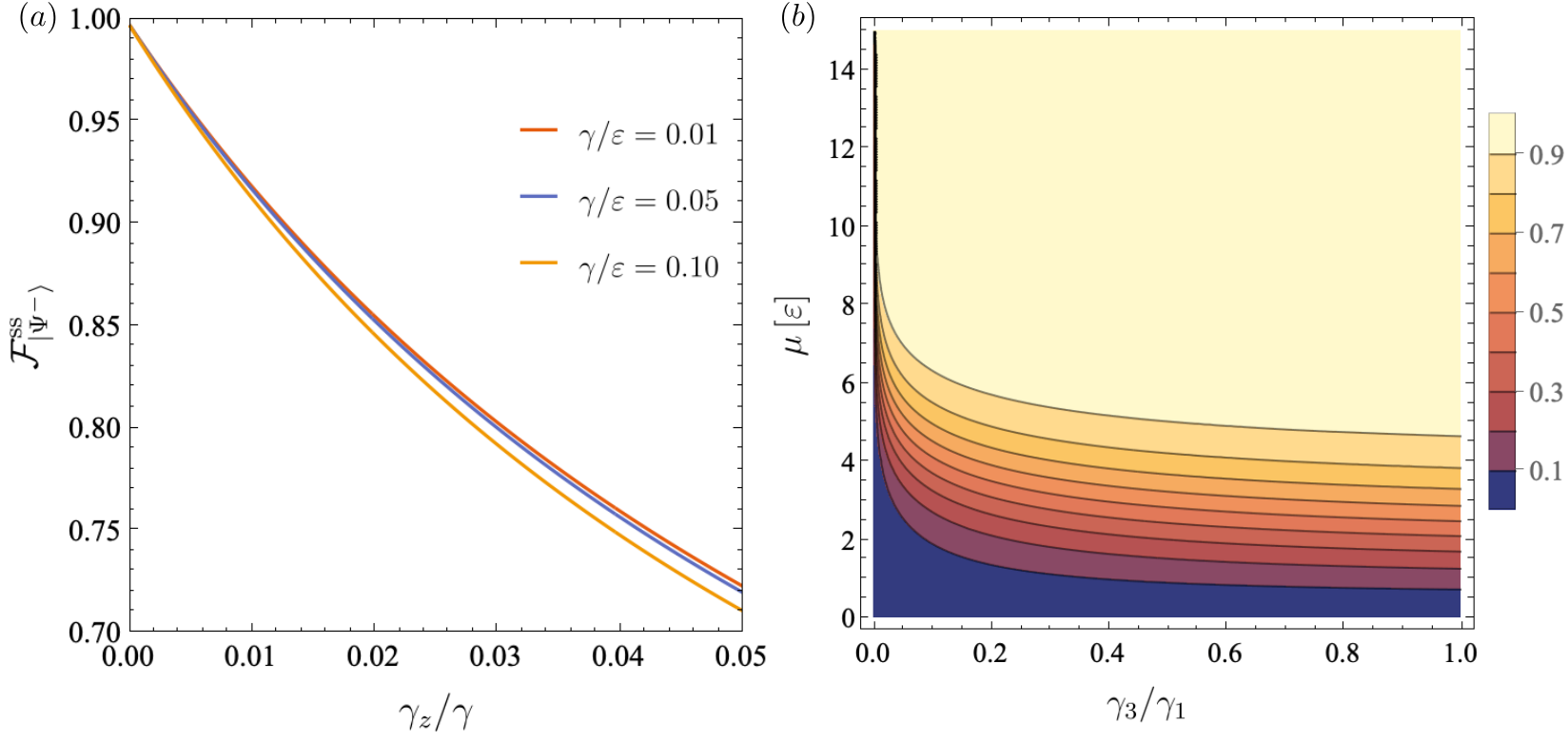}
	\caption{(a) The steady-state fidelity $\mathcal{F}^{\text{ss}}_{\ket{\Psi^-}}$  as a function of the ratio $\gamma_z/\gamma$ with $\mu/\varepsilon=8$. (b) The steady-state fidelity $\mathcal{F}^{\text{ss}}_{\ket{\Psi^-}}$  as a function $\mu/\varepsilon$ and the ratio of the couplings $\gamma_3/\gamma_1$.  The other parameters are $\varepsilon=1$, $T/\varepsilon=1$, $g_{23}/\varepsilon=g_{13}/\varepsilon=0.05$, $\mu/\varepsilon=8$, $U/\varepsilon=25$. }
	\label{fig:dephasing}
\end{figure}

\section*{D. Results beyond the Lindblad equation}	
In this section, we test the robustness of our findings beyond the Lindblad formalism considered throughout our work. There are two main issues to be addressed: (i) The analytical results presented rely on the three qubits being resonant in energy. A detuning in them may affect the quality of the entanglement that can be created. This is particularly relevant as our formalism ignores the Lamb-shift, which may shift the energy-resonance condition. Moreover, the local Lindblad equation does not hold for large values of the detuning (i.e., when the detuning is comparable to the energy gaps themselves \cite{Potts2021}) (ii) The Lindblad equation, being restricted to the lowest order in the system-reservoir couplings naturally ignores higher-order processes which can lower the generated entanglement.  To address these questions, we perform additional calculations with the second-order von Neumann (2vN) approach \cite{Pedersen2005,Pedersen2007}. This approach treats all system-reservoir couplings up to the second order, as well a class of higher order terms within a self-consistent treatment. 2vN has been shown to reproduce co-tunneling (i.e. the process of subsequent environment transitions where the intermediate state is energetically not allowed for). Such effects are highly relevant for the lifetime of the singlet state, which has no direct processes for its decay when restricting to the Lindblad formalism. The following simulations have been obtained with the QMEQ package \cite{KirsanskasComputPhysCommun2017} (the code is open-source \cite{github}). 

We found that the results are less sensitive to perturbation in $\varepsilon_3$, so we focus on the more interesting case $\varepsilon_1\neq\varepsilon_2$, setting $\varepsilon_1=\varepsilon_3=\varepsilon=1$ and varying $\varepsilon_2$.  As a reference point we used the parameters from the centre of Fig.~3 from the main text ($U/\varepsilon=15$ and $\mu/\varepsilon=8$), where the concurrence was very close to one. In Fig. \ref{fig:detuning} (a), we find that the Lindblad calculation for $g/\varepsilon=0.1$ yields concurrence greater than 0.95 for $0.98\lesssim\varepsilon_2/\varepsilon\lesssim 1.02$, i.e., for almost 2\% detuning.  For $g/\varepsilon=0.2$, this range is doubled to more than 4\%. The robustness further increases for $g/\varepsilon=0.5$.  Therefore, while ideally we want resonance between the qubit energies, both approaches show high concurrence in the presence of small fluctuations. We further examine the case in which the inter-qubit coupling becomes comparable to the energy of the dots, $g\gtrsim \varepsilon $, where the local Lindblad equation is no longer valid \cite{Potts2021}. Here, as expected, we find that there is a considerable mismatch between the the more accurate 2vN approach and the local Lindblad description that we have used in the manuscript. Moreover, there is considerable loss of concurrence even for zero detuning. This is expected, as the Bell state is no longer the steady state of the dynamics. In the case of large inter-qubit coupling, one may rely on a global master equation \cite{Potts2021} to find the true steady state.

In Fig. \ref{fig:detuning} (b), the role of the system-reservoir coupling $\gamma$ is examined. The deviation between both approaches shows two different features: With increasing $\gamma$ the peak position is shifted to larger values of $\varepsilon_2$ and the peak value is reduced. This is further analysed in Fig. \ref{fig:detuning} (c), where we use an even larger value of $\gamma = 0.3\varepsilon$ and add the result from the first-order von Neumann (1vN) approach [10], which shows the shift in peak position, but not the reduced peak value. As the 1vN (similar to the Redfield equations, which provide indiscernible results) restricts to system-reservoir couplings up to the first order, we can attribute the reduction of the peak value to higher-order effects, such as co-tunneling events, providing a path to empty the singlet state $\ket{\Psi^-}$. On the other hand, the peak shift arises in the 1vN approach due to the presence of Lamb-shift terms related to principle value integrals with the bath occupation functions. We explicitly checked, that neglecting these terms in the 1vN approach reproduces essentially the results from our Lindblad approach (which disregards such terms).
\par
Overall, it is clear from our analysis that we would like to remain well within the validity of the Lindblad approach, where the peak concurrence is close to one and the Lamb shift is negligible. For this, $\gamma\leq 0.1T = 0.1\varepsilon$ is a reasonable bound with deviations up to those visible in Fig. \ref{fig:detuning} (a).

\begin{figure}
	\centering
	\includegraphics[width=1.0\textwidth]{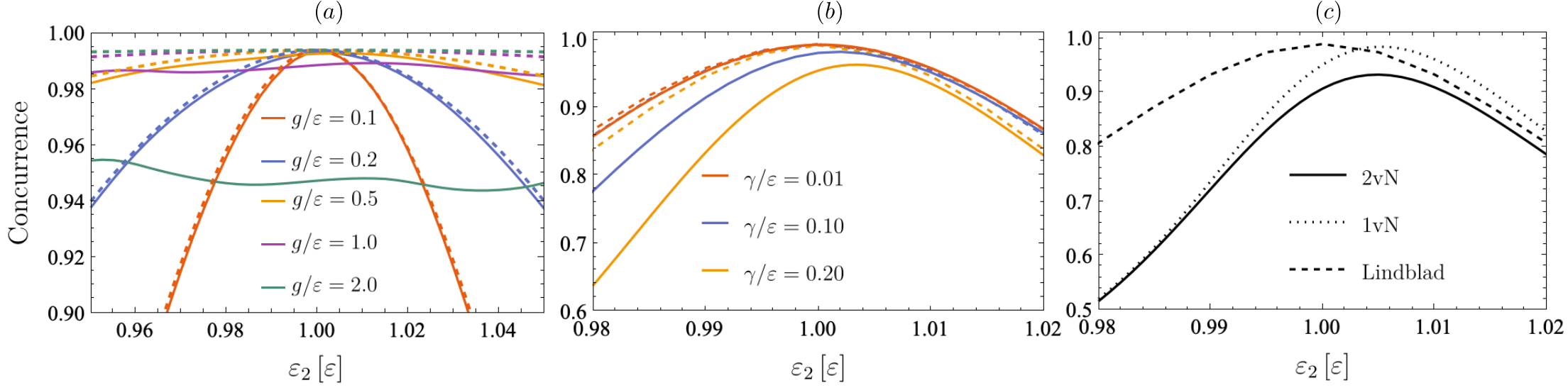}
	\caption{Concurrence as a function of the energy of the second qubit $\varepsilon_2$, with $\varepsilon_1=\varepsilon_3=\varepsilon=1$ and $T_1=T_3=T=1$. The inter-qubit couplings $g_{23}=g_{13}=g$ and system reservoir couplings $\gamma_1=\gamma_3=\gamma$.  (a) Concurrence as a function of $\varepsilon_2$ with constant $\gamma/\varepsilon=0.01$ and five different values of $g$, with the Lindblad and 2vN approaches. (b) The same with constant $g/\varepsilon=0.05$ and three values of $\gamma$. (c) Concurrence as a function of $\varepsilon_2$, with $g/\varepsilon=0.05$ and relatively large system-reservoir coupling $\gamma/\varepsilon=0.3$, with Lindblad, 2vN and 1vN approaches. In all three plots, the solid curves correspond to 2vN results and the dashed ones to Lindblad results. In panel (c), the dotted curve corresponds to a 1vN calculation. The other parameters are $U/T=15$, $\mu/T=8$.
	}\label{fig:detuning}
\end{figure}

\end{document}